\documentclass[%
 reprint,
preprintnumbers,
nofootinbib,
 amsmath,amssymb,
 aps,
]{revtex4-2}

\usepackage{graphicx}
\usepackage{dcolumn}
\usepackage{bm}
\usepackage{hyperref}
\usepackage{ulem}
\usepackage{multirow}

\DeclareUnicodeCharacter{2212}{\textendash}
\def\be{\begin{equation}}
\def\ee{\end{equation}}
\def\bea{\begin{eqnarray}}
\def\eea{\end{eqnarray}}

\usepackage{comment}
\renewcommand{\comment}[1]{}

\newcommand{\s}{\sigma}

\renewcommand{\d}{\delta}

\newcommand\sparbox[3][c]{\parbox[#1]{#2}{\strut#3\strut}}

\newcommand{\vs}{\nonumber\\}

\usepackage{xcolor}

\makeatletter
\newcommand{\Biggg}{\bBigg@{3.5}}
\makeatother

\definecolor{DarkGreen}{rgb}{.55,.71,0}

\begin{document}

\preprint{TUM-HEP-1507/24}

\title{Hot New Early Dark Energy bridging cosmic gaps: Supercooled phase transition reconciles (stepped) dark radiation solutions to the Hubble tension with BBN}

\author{Mathias Garny$^1$}\email{mathias.garny@tum.de}
\author{Florian Niedermann${}^2$}\email{florian.niedermann@su.se}
\author{Henrique Rubira$^1$}\email{henrique.rubira@tum.de}
\author{Martin S. Sloth${}^3$}\email{sloth@sdu.dk}
\affiliation{%
\small $^1$ Physik Department T31, Technische Universit\"at M\"unchen \\
James-Franck-Stra\ss e 1, D-85748 Garching, Germany\\
${}^2$ Nordita, KTH Royal Institute of Technology and Stockholm University\\
Hannes Alfv\'ens v\"ag 12, SE-106 91 Stockholm, Sweden
\\
${}^3$ Universe-Origins, University of Southern Denmark, Campusvej 55, 5230 Odense M, Denmark}%

\begin{abstract}
We propose a simple model that can alleviate the $H_0$ tension while remaining consistent with big bang nucleosynthesis (BBN). It is based on a dark sector described by a standard Lagrangian featuring a $SU(N)$ gauge symmetry with $N\geq3$ and a massive scalar field with a quartic coupling. The scalar acts as dark Higgs leading to spontaneous symmetry breaking $SU(N)\to SU(N\!-\!1)$ via a first-order phase transition \`a la Coleman-Weinberg. This set-up naturally realizes previously proposed scenarios featuring strongly interacting dark radiation (SIDR) with a mass threshold within hot new early dark energy (NEDE). For a wide range of reasonable model parameters, the phase transition occurs between the BBN and recombination epochs and releases a sufficient amount of latent heat such that the model easily respects bounds on extra radiation during BBN while featuring a sufficient SIDR density around recombination for increasing the value of $H_0$ inferred from the cosmic microwave background. Our model can be summarized as a natural mechanism providing two successive increases in the effective number of relativistic degrees of freedom after BBN but before recombination $\Delta N_\mathrm{BBN} \to \Delta N_\mathrm{NEDE} \to \Delta N_\mathrm{IR}$ alleviating the Hubble tension. The first step is related to the phase transition and the second to the dark Higgs becoming non-relativistic. This set-up predicts further signatures, including a stochastic gravitational wave background and features in the matter power spectrum that can be searched for with future pulsar timing and  Lyman-$\alpha$ forest measurements.
\end{abstract}

\maketitle


\section{Introduction}\label{introduction}

The Hubble tension has risen to become one of cosmology's most hotly debated problems \cite{Abdalla:2022yfr} and persists despite increasing levels of scrutiny~\cite{Riess:2024ohe}. Assuming it is not due to still unaccounted systematics in the measurements, a resolution of  the Hubble tension will have to involve new physics, going beyond the $\Lambda$CDM model, at relatively low ($\sim$\,eV) energy scales, where it will affect the CMB and other precision probes of the evolution of the Universe \cite{Bernal:2016gxb,Knox:2019rjx}. This makes the possible scenarios for resolving the Hubble tension very constrained and also very testable.

One of the earliest and simplest proposed solutions to the Hubble tension is the existence of a hypothetical fluid of Strongly Interacting Dark Radiation (SIDR) \cite{Buen-Abad:2015ova,Buen-Abad:2017gxg,Archidiacono:2020yey}. The SIDR solution has several incarnations. One is the so-called ``stepped" model, where the SIDR fluid temporarily becomes non-relativistic and then decays into a relativistic fluid again, creating a small step in the number of effective relativistic degrees of freedom, $N_{\textrm{eff}}$ \cite{Aloni:2021eaq,Schoneberg:2022grr,Buen-Abad:2022kgf,Allali:2023zbi}.  This mechanism allows for an $\ell$-dependence in the CMB, with different impacts of the SIDR fluid on the high-$\ell$ and low-$\ell$ multipoles. 

A common problem of the SIDR model, including the stepped models, is that they are ruled out by BBN as solutions to the Hubble tension when taken at face value. Specifically, to address the Hubble tension, they require an initial value of the effective number of relativistic degrees of freedom, $N_{\textrm{eff}}$, which is too large to be compatible with BBN constraints~\cite {Schoneberg:2022grr}. 
This problem is usually circumvented by arguing that ``something'' could happen after BBN, creating the initially large $N_{\textrm{eff}}$ required in the SIDR and the stepped models, but this ``something'' has not been included in the models in the literature so far at the detailed level (although for some initial suggestions see \cite{Berlin:2017ftj,Berbig:2020wve,Escudero:2022gez,Aloni:2023tff} and Sec.~\ref{steps} below). Without further modification, SIDR and stepped models are therefore ruled out as solutions to the Hubble tension by BBN constraints.   

In fact, we take the point of view that BBN physics can not be separated from the solution to the Hubble tension when fitting the CMB. This is emphasized by the fact that the details of CMB and recombination physics depend on the helium fraction predicted by BBN.
Therefore, the CMB and recombination physics are not independent from BBN physics.\footnote{Assuming a different value of $\Delta N_\mathrm{eff}$ at the time of BBN than around the recombination era would require manually setting the helium abundance at the CMB epoch to an assumed value. Instead, having a complete model allows one to consistently infer the helium abundance from BBN.} The predictions of BBN are important for the full subsequent thermal evolution of the Universe. Therefore, a serious complete model solving the Hubble tension must also be required to be consistent with BBN constraints. 

In this work we show how, in terms of a dark sector described by a simple microphysical model, a first-order phase transition triggered by the temperature of the dark sector
(named Hot New Early Dark Energy [Hot NEDE] in~\cite{Niedermann:2021vgd,Niedermann:2021ijp}) provides the UV completion of the SIDR and stepped models, bringing them in agreement with the BBN constraints as solutions to the Hubble tension. 
In particular, we consider a standard Lagrangian described by an $SU(N)$ gauge symmetry and a massive dark Higgs field with quartic coupling as only ingredients, leading to  spontaneous symmetry breaking $SU(N)\to SU(N\!-\!1)$. As pointed out by Witten~\cite{Witten:1980ez}, this set-up  features a strongly supercooled first-order phase transition  \`a la Coleman-Weinberg~\cite{Coleman:1973jx} in the conformal limit where the effective scalar mass vanishes. In conclusion, when the latent heat of the false vacuum is converted into light species in the phase transition, the number of effective relativistic degrees of freedom, $N_{\textrm{eff}}$, is suddenly increased. 

If the Hot NEDE phase transition occurs between the BBN and recombination epochs, i.e. roughly within the wide redshift range $10^5\lesssim z_\ast< 10^9$, this explains how a relatively large $N_{\textrm{eff}}$, as required in the SIDR and stepped models, can be made consistent with BBN constraints. The dark radiation component generated after the phase transition naturally gives rise to a sizeable change $\Delta N_{\textrm{eff}}\sim {\cal O}(1)$. It is composed of massless $SU(N\!-\!1)$ gauge bosons  as well as a light dark Higgs boson, featuring self-interactions as assumed in SIDR models generated by the remaining non-Abelian $SU(N\!-\!1)$ gauge interactions for $N\geq 3$. In addition, a small mass of the dark Higgs is a characteristic feature of the Coleman-Weinberg mechanism. Thus, it turns non-relativistic somewhat after the phase transition, leading to a further slight increase of $N_{\rm eff}$ in a second ``step'', and providing a natural realization of the ``stepped'' SIDR framework.

In Sec.~\ref{sec:prev} we will review the NEDE framework and the SIDR models as solutions to the Hubble tension. In Sec.~\ref{origin} we explain our new dark sector model. In Sec.~\ref{data} we compare our model with cosmological datasets and discuss how it can alleviate the Hubble tension. Then in Sec.~\ref{sec:pheno} we discuss novel phenomenological signatures of the model, and finally we conclude and give pointers for future work in Sec.~\ref{sec:conslusion}.  The appendices contain further details on our results, as well as equations used to describe the evolution after the phase transition.

\section{Background and preliminaries}\label{sec:prev}

\subsection{Comparison to Cold NEDE}

NEDE is a framework for addressing the Hubble tension in terms of a new phase of dark energy which decays in a fast-triggered phase transition before or around matter-radiation equality \cite{Niedermann:2019olb,Niedermann:2020dwg,Niedermann:2023ssr}. It is distinct from other early dark energy models, most notably \mbox{axiEDE}~\cite{Poulin:2018cxd} (for a review see~\cite{Poulin:2023lkg}) both in terms of its detailed phenomenology and field-theoretic realization.  

Hot NEDE \cite{Niedermann:2021vgd,Niedermann:2021ijp, Cruz:2023lnq} is also different from the previously studied Cold NEDE models \cite{Niedermann:2019olb,Niedermann:2020dwg}, in which the phase transition is triggered by an Ultra-Light Axion-like (ULA) scalar field at zero dark sector temperature \cite{Cruz:2023lmn}. In particular, the Cold NEDE fluid is usually assumed to have a constant equation of state $w \simeq 2/3$ after the phase transition as a phenomenological requirement for solving the Hubble tension. We note, however, that $N_{\textrm{eff}}$ is the same in Cold NEDE and in $\Lambda$CDM, making it trivially consistent with BBN constraints. In Hot NEDE, there is no ULA, and the phase transition is instead triggered by the dark sector temperature corrections to the NEDE boson's effective potential. After the phase transition, the latent heat of the false vacuum goes into reheating the dark sector, creating a non-trivial change in $\Delta N_{\textrm{eff}}$.

One of the main attraction points of Hot NEDE as a solution to the Hubble tension is that it is a phase transition similar to other transitions such as the QCD and electroweak transitions that are already part of our understanding of the thermal history of the universe. As we point out,  the microscopic description of Hot NEDE  can  be reminiscent of a dark version of the electroweak phase transition with a light Higgs. Besides, it could also be related to other open questions, such as the origin of neutrino masses \cite{Niedermann:2021vgd,Niedermann:2021ijp}.  However, while Cold NEDE has been implemented into a Boltzmann code and studied rigorously in the past, a phenomenological test of Hot NEDE has not yet been performed. In this paper, we will, for the first time, implement Hot NEDE into a Boltzmann code and test it against data. 

Previously, it was assumed that the phenomenology of Hot NEDE from a CMB point of view would be very similar to Cold NEDE. In this work, however, we point out that the phenomenology of Hot and Cold NEDE differs greatly. The fluctuations in the thermal trigger of Hot NEDE are larger than in the scalar trigger of Cold NEDE. This means that the Hot NEDE model cannot resolve the Hubble tension by simply decaying into a semi-stiff fluid with the equation of state $w \simeq  2/3$ around redshift $z_\ast=5000$, as is typical in previously studied (N)EDE models, like Cold NEDE. Instead, in Hot NEDE, the phase transition must occur earlier (but after BBN) at a redshift $10^5 \lesssim z_\ast <  z_{\textrm{BBN}}\sim  10^{9}$.  After the phase transition, the NEDE boson and gauge fields will naturally behave like a SIDR fluid, mimicking the SIDR and stepped models around recombination yet with a natural explanation for the initial high  $N_{\textrm{eff}}$ created after BBN, and in this way UV completing the SIDR and stepped models as solutions to the Hubble tension.

\subsection{Previous steps attempted}\label{steps}

As exemplified by the initial suggestions mentioned above \cite{Berlin:2017ftj,Berbig:2020wve,Escudero:2022gez,Aloni:2023tff}, increasing $N_\text{eff}$ after BBN is not easy. One obvious idea would be to have an extra mass threshold leading to a second step in the stepped SIDR model, with both steps in $N_\mathrm{eff}$ happening after BBN. Another of the previously discussed possibilities is having some dark degrees of freedom which thermalize late through the neutrino sector after BBN. Let us quickly discuss both of these ideas for generating a sufficiently large $N_\mathrm{eff}$ after BBN, before we proceed to discuss in detail how a phase transition with a certain amount of supercooling between BBN and before recombination provides a conceptually simple mechanism.

As mentioned, one may wonder whether a ``double step'' scenario could reconcile the extra radiation needed to address the $H_0$ tension within the stepped SIDR model with BBN constraints. This would be a scenario with a dark sector featuring one mass threshold to address $H_0$ and an earlier one after BBN to obtain a sufficient energy density in the dark sector. Assume (for simplicity) real scalars, with $n_{\text{IR}}<n_{\text{UV}}<n_{\text{BBN}}$ degrees of freedom, where ``UV'' refers to the period between the two steps, BBN before them, and ``IR'' after them. The contribution to $N_\text{eff}$ is then $\Delta  N_{\text{IR}}=(n_{\text{UV}}/n_{\text{IR}})^{1/3} \Delta  N_{\text{UV}}$
and $\Delta  N_{\text{UV}}=(n_{\text{BBN}}/n_{\text{UV}})^{1/3} \Delta  N_{\text{BBN}}$.
Given BBN bounds $\Delta N_{\textrm{eff}}=-0.11\pm0.23$~\cite{Yeh:2022heq} or $\Delta N_{\textrm{eff}}=-0.10\pm0.21$~\cite{schöneberg20242024} we require $\Delta  N_{\text{BBN}}\lesssim 0.1$ in order not to introduce any tension with BBN. Assuming the dark sector was in thermal contact in the Early Universe, the minimal contribution to $\Delta N_{\textrm{eff}}$ for a real scalar is $0.027$ (e.g.~\cite{Baumann:2017gkg}).
This means we can have at most $n_{\text{BBN}}= 4$. Further assuming the minimal possible $n_{\text{IR}}=1$
and $n_{\text{UV}}=2$ this means we can have at most $\Delta  N_{\text{UV}}=(4/2)^{1/3}4\times 0.027\simeq 0.14$ and $\Delta  N_{\text{IR}}=(2/1)^{1/3} \Delta  N_{\text{UV}}\simeq 0.17$, which is too low to address the $H_0$ tension. To achieve the desired $\Delta  N_{\text{IR}}\simeq 0.6$ one would, even in the most favourable case with $n_{\text{IR}}=1$, need $n_{\text{BBN}}=(0.6/0.027)^{3/4}\simeq 10$ degrees of freedom during BBN, which would mean $\Delta  N_{\text{BBN}}\simeq 0.27$. Although not being completely excluded, this would arguably just trade the $H_0$ for a BBN tension.

A logical possibility for which the ``double step''  could work in principle is to assume that the dark sector was never in thermal equilibrium with the SM and populated with a lower temperature. While this by itself is a plausible possibility, enhancing the dark sector density sufficiently just due to mass threshold effects is still challenging in this scenario. Specifically, to have $\Delta  N_{\text{BBN}}\lesssim 0.1$ but nevertheless achieve $\Delta  N_{\text{IR}}\simeq 0.6$ via mass threshold effects for addressing the $H_0$ tension would require a large number of particles in the dark sector with $n_{\text{BBN}}\simeq (0.6/0.1)^3 ~n_{\text{IR}}\simeq 200~n_{\text{IR}}\geq 200$ degrees of freedom (where we assumed the minimal value $n_{\text{IR}}=1$ in the last estimate) that all exhibit a mass threshold after BBN but sufficiently before recombination~\cite{Aloni:2023tff}. Having more degrees of freedom than contained in the entire Standard Model (SM) that undergo a non-trivial thermal evolution in the dark sector in the energy range between BBN and recombination is, however, a rather non-minimal proposal.

It has also been suggested that the constraints from BBN could be avoided if the dark sector temperature was effectively zero at the time of BBN, and only created after BBN through thermalization of dark sector fermions with the SM neutrinos, which have decoupled at this point~\cite{Aloni:2023tff}. Since the SM neutrinos have decoupled from the SM, the thermalization with the dark sector does not by itself change $N_\mathrm{eff}$, as it does not change the total energy density of the dark sector and SM neutrinos. However, if one or more of the self-interacting dark fermions subsequently undergo a mass threshold and annihilate into a much lighter force carrier, then that would lead to a second step as discussed above, but without affecting BBN if the dark fermions have a mass $100$ eV $\lesssim m_{\nu_d} \lesssim 100$ keV. In \cite{Aloni:2023tff} it was argued that this allows for $\Delta N_\mathrm{eff} =  [(g_\mathrm{rel, d}^\mathrm{UV}/g_\mathrm{rel, d}^\mathrm{IR})^{1/3}-1]N_\mathrm{eq}$ where $N_\mathrm{eq}$ is the number of SM neutrinos equilibrating with the dark fermions, and $g_\mathrm{rel, d}^\mathrm{UV}$ and $g_\mathrm{rel, d}^\mathrm{IR}$ are the effective number of degrees of freedom before and after the mass threshold respectively. To have $\Delta N_\mathrm{eff}\approx 0.6$, as relevant for solving the Hubble tension, one would need four dark self-interacting fermions with masses $100$ eV $\lesssim m_{\nu_d} \lesssim 100$ keV equilibrating with the SM neutrinos through a non-vanishing mixing angle $\theta > 10^{-13}$. However, it is plausible that the SIDR after the mass threshold consists of more than a single degree of freedom, making $g_\mathrm{rel, d}^\mathrm{IR}>1$, and hence a very large dark sector equilibrating with the SM neutrinos after BBN is again quickly required.\footnote{In addition the proposed mixing of active neutrinos and a sterile neutrino with a self-interaction mediated by a light boson also suffers model-dependent constraints from stellar and supernova cooling \cite{Archidiacono:2014nda,Li:2023vpv}.}

While this remains a viable possibility, we will instead consider a simple new proposal, namely that both a significant dark sector temperature and a $\Delta N_\mathrm{eff}$ are created by an energy injection into the dark sector from a super-cooled phase transition between BBN and recombination in the Hot NEDE scenario. As we will see, all desired ingredients to realize this scenario naturally emerge from a dark sector described by a standard set-up from the point of view of particle physics.

\section{Origin of dark species}\label{origin}

\begin{figure*}[t]
	\begin{center}
	   \includegraphics[width=0.95\textwidth]{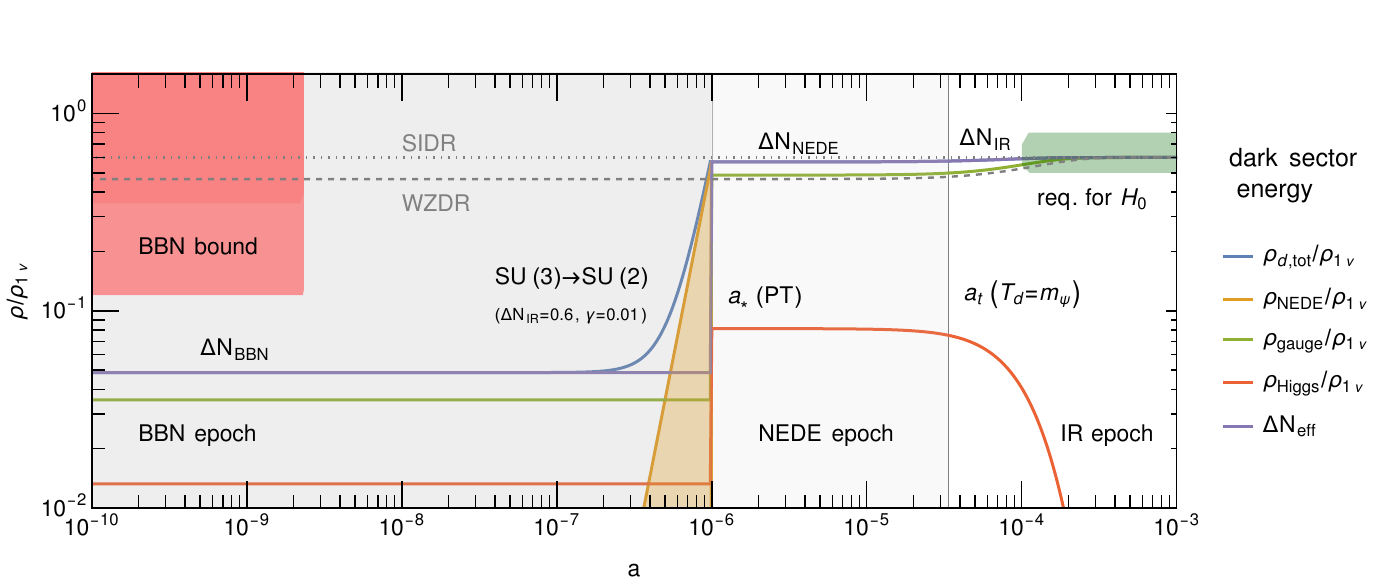}
    \end{center}
    \caption{Energy density of different dark sector components in units of $\rho_
{1,\nu} \equiv \frac{7}{4}\frac{\pi^2}{30} \left(\frac{4}{11}\right)^{4/3}T_{\rm vis}^4$. The blue line depicts the evolution of the total dark sector energy density, composed of the dark radiation plasma formed by the gauge (green line) and Higgs (red line) bosons as well as the latent heat $\rho_{\rm NEDE}$ (orange line and area), that rises above the radiation fluid before the supercooled phase transition. The purple line depicts $\Delta N_{\rm eff}$  before and after the phase transition at $a_*$, respectively, see \eqref{eq:rho_before}  and \eqref{eq:rho_after}. The red shaded area illustrates BBN constraints, and the green area values of $\Delta N_{\rm eff}$ required for addressing the $H_0$ tension. For comparison grey dotted and dashed lines show the SIDR model and its stepped version named WZDR~\cite{Aloni:2021eaq} (see Sec.~\ref{introduction}).}\label{fig:dsenergy}
\end{figure*}

\begin{table*}[t!h]
\renewcommand{\arraystretch}{1.3}
	\centering
	\begin{tabular}{|c||c|c|c|} \hline 
		\bf{Parameter} & \bf{Eq.}& \bf{Reference model}  & \bf{Description}    \\ 
    \hline \hline
		$g$           & \eqref{eq:model}      &$0.05$                 & \sparbox{9cm}{Gauge coupling with lower bound [see~\eqref{eq:Gamma_2}] $g\gtrsim 0.01$.} \\ \hline 
		$N$           & \eqref{eq:model}      &$3$                    & \sparbox{9cm}{Gauge group breaking pattern: $SU(N)\to SU(N\!-\!1)$.} \\ \hline
  		$v$ & \eqref{Veff}   &$1.4 \times 10^4 \, \mathrm{eV}$&   \sparbox{9cm}{NEDE field vacuum expectation value ({\rm vev}) arising from dimensional transmutation.} \\ \hline
		$\mu_{\rm eff}$         & \eqref{Veff}    &$1.1 \, \mathrm{eV}$ & \sparbox{9cm}{ Renormalized mass parameter of NEDE field, describes small soft breaking of classical conformal symmetry for $\mu_{\rm eff}\ll g^2v$, provides  graceful exit from supercooling.} \\  
      \hline\hline 
        $m_A \equiv g v /2$& --               &$3.5  \times 10^2 \, \mathrm{eV}$  &   \sparbox{9cm}{Gauge boson mass scale after symmetry breaking.} \\ \hline
  		$m_{\psi} \sim g^2v$ &    \eqref{Higgsmass}        &$4.4 \, \mathrm{eV}$&   \sparbox{9cm}{Dark Higgs mass.} \\ \hline
  		$\Delta V_* \sim g^4v^4$ &    \eqref{def:latent_heat}       &$(127 \, \mathrm{eV})^4$&   \sparbox{9cm}{Latent heat released during the phase transition.} \\ \hline
		   $\lambda\sim g^4$     &  \eqref{eq:vac_pot}   & \sparbox{3cm}{$5.0 \times 10^{-8}$} & \sparbox{9cm}{Tree-level self-coupling of the NEDE field.} \\\hline 
			$\gamma\sim \mu_{\rm eff}^2/(g^4v^2)$      &\eqref{def:gamma}      &$0.01 $               & \sparbox{9cm}{Supercooling parameter, strong supercooling for $\gamma \ll 1$.} \\ 
  \hline \hline
  		$T_c\sim gv$     &\eqref{Tc} &$1.2 \times 10^2 \, \mathrm{eV}$&  \sparbox{9cm}{Dark sector temperature with two degenerate vacua.} \\ \hline
        $T^\ast_d\sim \sqrt{\gamma}gv$     &\eqref{xi_star} & $39.2 \, \mathrm{eV}$ &  \sparbox{9cm}{Dark sector temperature when percolation condition is met ($T_d^* \simeq T_b $ in the supercooling regime).} \\ \hline
        $T_b\sim \sqrt{\gamma}gv$     & \eqref{Tb} & $39.1 \, \mathrm{eV}$& \sparbox{9cm}{Dark sector temperature below which thermal barrier vanishes.} \\ \hline
        $T^\mathrm{*,after}_d\sim\Delta V_*^{1/4}$     &\eqref{eq:xi_after} & $ 1.0 \times 10^2 \, \mathrm{eV}$ &  \sparbox{9cm}{Dark sector temperature after the phase transition.} \\ \hline
        $T^*_\mathrm{vis}$     &\eqref{xi_star} & $ 2.4 \times 10^2 \, \mathrm{eV}$ &  \sparbox{9cm}{Visible sector temperature during the phase transition.} \\ 
        \hline
        $H_*/\beta$     & \eqref{eq:beta} & $ 2.9 \times 10^{-6}$ &  \sparbox{9cm}{Percolation time scale.} \\         \hline
        $\alpha$     & -- & 0.076 &  \sparbox{9cm}{Strength of phase transition, $\alpha \equiv \Delta V /( \rho_\mathrm{rad,vis}^* + \rho_\mathrm{rad,d}^*)$.} \\         \hline
	 \hline
  		$z_\ast$  &-- &$10^6$& \sparbox{7cm}{Redshift of the NEDE phase transition (first step) with possible range: $\left[ 10^{5}, 10^{9}\right]$.} \\ \hline
      	$z_t\sim gz_\ast$    &\eqref{def:z_t}& $4.2 \times 10^4$& \sparbox{9cm}{Redshift when the dark Higgs becomes non-relativistic, i.e.~for which $m_\psi = T_d$ (second step) with $z_t < z_\ast$.} \\ \hline
		$f_{\rm NEDE}$     & \eqref{def:latent_heat}&$0.071$& \sparbox{9cm}{Fraction of latent heat at redshift $z_\ast$, setting the size of the first step. Note that $f_{\rm NEDE}=\alpha/(1+\alpha)$.} \\ \hline
    		$r_g$     & \eqref{def:rg}&1/6 & \sparbox{7cm}{Size of the second step.} \\ \hline\hline

		$\Delta N_{\rm BBN}$     &\eqref{N_BBN} &0.039&  \sparbox{9cm}{Contribution to $\Delta N_\mathrm{eff}$ before the NEDE phase transition, applicable during BBN.} \\ \hline
		$\Delta N_{\rm NEDE}$   & \eqref{eq:N_NEDE} & 0.57&  \sparbox{9cm}{Contribution to $\Delta N_\mathrm{eff}$ after the NEDE phase transition.} \\ \hline
		$\Delta N_{\rm IR}$   &\eqref{eq:N_IR} &0.6 &  \sparbox{9cm}{Contribution to $\Delta N_\mathrm{eff}$ after the second step.} \\ 
		\hline
	\end{tabular}
	\caption{Summary of the fundamental and phenomenological model parameters. The reference values correspond to a typical Hot NEDE cosmology (also depicted in Fig~\ref{fig:dsenergy}). Once $\mu_\mathrm{eff}$, $v$, $g$, and $N$, alongside an initial condition for $\Delta N_\mathrm{BBN}$, are fixed, all other parameters can be derived.}

	\label{table:energy_scales}
\end{table*}

Our dark sector model relies on a local $SU(N)$ gauge symmetry, which is spontaneously broken to  $SU(N\!-\!1)$ by a scalar field undergoing a first-order phase transition. Our main focus centers on the evolution of this sector's contribution to the effective number of relativistic degrees of freedom, $\Delta N_\mathrm{eff}$, before and after the phase transition, featuring an increase related to the latent heat (the first step). Later, yet less importantly, the scalar field driving the phase transition becomes non-relativistic, leading to a second change in  $N_\mathrm{eff}$ (the second step), in a process similar to the step described in~\cite{Aloni:2021eaq}. The evolution of $\Delta N_\mathrm{eff}$ as well as the various dark sector energy densities are shown in Fig.~\ref{fig:dsenergy} for a typical set-up and all relevant parameters are summarized in Tab.~\ref{table:energy_scales}. We discuss how this conceptually simple and well-known set-up naturally generates a consistent cosmological evolution throughout BBN, recombination and all the way until the present era, alleviating the Hubble tension while retaining the success of BBN.

We introduce the model's Lagrangian in Sec.~\ref{subsec:model}. Then, we derive the parameter regime corresponding to a supercooled phase transition in Sec.~\ref{subsec:pt}. The latent heat deposited in the dark sector leads to a very quick reheating as discussed in Sec.~\ref{subsec:reheating} This is followed by a review of the second step in Sec.~\ref{subsec:2ndstep}. Finally, in Sec.~\ref{subsec:perts} we describe the perturbations in the plasma.

\subsection{The model}
\label{subsec:model}
We consider a dark sector featuring a gauge symmetry $SU(N)$ with gauge coupling $g$ and a complex scalar field $\Psi$ described by the standard Lagrangian
\be\label{eq:model}
    \mathcal{L} = |D\Psi|^2 - V_{\rm cl}(|\Psi|^2) -\frac12 \text{tr}\,F_{\mu\nu}F^{\mu\nu}  \,,
\ee
where $D$ is the gauge covariant derivative, $D^\mu \Psi = \partial^\mu \Psi - ig A_a^\mu \tau^a \Psi$ with generators satisfying $[\tau^a,\tau^b]=if^{abc}\tau^c$ with structure constants $f^{abc}$, and $F^{\mu\nu}=ig^{-1}[D^\mu,D^\nu]$ the field strength tensor. The dark Higgs field $\Psi$ transforms in the fundamental representation of $SU(N)$, and we refer to it as the NEDE scalar field. Its zero temperature tree-level potential is given by 
\be\label{eq:vac_pot}
V_{\rm cl}(|\Psi|^2) = -\mu^2 |\Psi|^2 + \lambda |\Psi|^4 + V_0\,, 
\ee
where $\mu^2$ is the field's (tachyonic) mass parameter and $\lambda$ its self-coupling. $V_0$ is an additive constant which is chosen such that the potential energy is zero in the true vacuum.

As we will discuss in detail later, the gauge field induces thermal corrections that will lead to a first-order phase transition at a redshift $z_\ast \equiv 1/a_\ast -1$ corresponding to a dark sector temperature $T_d^\ast$.\footnote{Here and henceforth, an asterisk is used as a shorthand for evaluation at the moment of the phase transition. If a quantity is discontinuous, we refer to its value \textit{before} the transition, e.g., $T^*_{d}$ is the (dark sector) temperature right before the phase transition (also sometimes denoted as $T_n$ in the literature).} 
As the symmetry breaking from  $SU(N) \to SU(N\!-\!1)$ occurs, the NEDE field picks up a vacuum expectation value  $v$. We parametrize the corresponding breaking pattern through 
        \be \label{eq:vev}
            \Psi = e^{2i\pi^a\tau^a/v}
        \begin{pmatrix}
	   0 \\ \dots \\
	   \frac{v}{\sqrt{2}} + \frac{ h_\psi}{\sqrt{2}}
        \end{pmatrix} \,,
        \ee
where $h_\psi$ is the physical Higgs mode of the NEDE field after the phase transition, and $\pi^a$ are the Goldstone fields associated with the $2N-1$ broken 
 generators $\tau^a$ that are absorbed by the gauge bosons acquiring mass. 
Let us consider the example of $SU(3)\to SU(2)$ in more detail. After the breaking there are three massless states, $m_{A_1} = m_{A_2} =  m_{A_3} = 0$, corresponding to $n_{A_{1-3}} = 6$ degrees of freedom. The remaining five gauge bosons are massive. Explicitly, $m_{A_4} = m_{A_5} =  m_{A_6} = m_{A_7} = \frac{gv}{2}$ 
with $n_{A_{4-7}} = 4\times3=12$, and  $m_{A_8} = \frac{gv}{\sqrt{3}}$ with  $n_{A_8} = 3$. The physical NEDE mode develops a vacuum mass $m_\psi$, of order $g^2v$, see~\eqref{Higgsmass} below. 
The number of massless and massive degrees of freedom in both phases is also summarized in Tab.\,\ref{tab:dof} for general $N$. 

\begin{table*}[t]
\renewcommand{\arraystretch}{1.3}
	\centering
	\begin{tabular}{|c||c|c|c|} \hline 
		\#dof & $z>z_\ast$ (``BBN'') & $z_\ast>z>z_t$ (``NEDE'') & $z<z_t$ (``IR'')  \\ \hline
    $g_{{\rm rel},d}$ & $2(N^2+N-1)$ & $2(N-1)^2-1$ & $2(N-1)^2-2$ \\ \hline
  massless gauge bosons & $2(N^2-1)$ & \multicolumn{2}{c|}{$2((N-1)^2-1)$} \\ \hline
  massive gauge bosons & $0$ & \multicolumn{2}{c|}{$n_A\equiv 3(N^2-(N-1)^2)=3(2N-1)$} \\ \hline
  Higgs bosons & $2N$ & \multicolumn{2}{c|}{$1$} \\ \hline
    \end{tabular}
    \caption{\label{tab:dof} Degrees of freedom (dof) in the dark sector. The first row shows the relativistic dof $g_{{\rm rel},d}$ before ($z>z_\ast$) and after ($z<z_\ast$) the $SU(N)\to SU(N\!-\!1)$ phase transition, further discriminating the regime where the Higgs is relativistic ($z>z_t$) or non-relativistic ($z<z_t$). The lower rows show the dof of each species. }
\end{table*}

\subsection{The phase transition} \label{subsec:pt}

We now describe the physics of the thermal phase transition at $z=z_\ast$, focusing on the 
regime of a supercooled first order phase transition, that permits a sizeable step in $\Delta N_\mathrm{eff}$. This occurs in the limit for which the dark Higgs mass is light compared to the gauge boson masses~\cite{Witten:1980ez}\footnote{See also \cite{Niedermann:2021ijp,Niedermann:2021vgd} for more details in the context of Hot NEDE, and e.g.~\cite{Yamamoto:1985rd,Barreiro:1996dx,Gildener:1976ih,Iso:2017uuu,Levi:2022bzt,Salvio:2023ynn,Kierkla:2023von,Freese:2023fcr} for a discussion of supercooling in other scenarios.}
\be \label{eq:masshierar}
	m_A \gg T_d^*, m_\psi\,,
\ee 
being realized for $\lambda\sim g^4$. Adopting this power counting, the phase transition can be described by the one-loop approximation of the effective potential taking gauge boson loops into account. It can be written as
\begin{align}\label{Veff}
V({\psi}; T_d) = & V_0-\frac{\mu_{\rm eff}^2}{2}{\psi}^2\left(1-\frac{\psi^2}{2v^2}\right)   + B\psi^4\left(\ln\frac{\psi^2}{v^2}-\frac12\right) \nonumber\\
& + \Delta V_\mathrm{thermal}({\psi}; T_d) \;,
\end{align}
where ${\psi} \equiv \sqrt{2}|\Psi|$ denotes the field value. Let us review the various contributions in turn. First of all, note that we traded the tree-level parameters $\mu^2$ and $\lambda$ in~\eqref{eq:model} by\footnote{The relation to \eqref{eq:model} is $\mu^2=\mu_{\rm eff}^2+4Bv^2=\lambda v^2$.} $\mu_{\rm eff}^2$ and $v^2$ such that $\mu_{\rm eff}^2=0$ corresponds to the classically conformal limit for which symmetry breaking famously occurs purely due to the radiative correction proportional to $B$ (see~\eqref{def:B}) and the scale $v$ is generated by dimensional transmutation as pointed out by Coleman and Weinberg~\cite{Coleman:1973jx}. For generality, we allow for non-zero (but potentially small) values $\mu_{\rm eff}^2$ of order up to $v^2g^4$, consistent with the power counting $\lambda\sim g^4$, which can be seen as a technically natural ``soft'' breaking of conformal symmetry. The parameter $B$ is 
\begin{equation}\label{def:B}
B=\frac{c_1 n_A}{64\pi^2}\left(\frac{g}{2}\right)^4\,,
\end{equation}
where $n_A$ is the number of gauge bosons acquiring mass (see Tab.~\ref{tab:dof}) and $c_1$ a constant of order unity ($c_1=52/45$ for $N=3$).\footnote{$c_1= \frac{3}{n_A}16\sum_a m_{A_a}^4(\psi)/(g\psi)^4$ and $c_0=\frac{3}{n_A}\sum_a m_{A_a}^2(\psi)/(g\psi)^2$.} In this parameterization the minimum of the vacuum potential in the first line of~\eqref{Veff} occurs at $\psi=v$, and the constant $V_0$ ensures $V(v;0)=0$.

A well-known case featuring very strong supercooling, discussed by Witten~\cite{Witten:1980ez} (see also \cite{Guth:1982pn}), is the conformal limit $\mu_{\rm eff}=0$. As we will see a small non-zero $\mu_{\rm eff}$ (related to the supercooling parameter $\gamma$ defined in~\eqref{def:gamma} below) provides a graceful exit mechanism allowing for a controlled amount of supercooling, and preventing the latent heat from dominating the total energy budget~\cite{Niedermann:2021vgd,Levi:2022bzt}. As in the purely conformal limit, the dark Higgs is parametrically lighter than the gauge boson masses,
\begin{equation}\label{Higgsmass}
m_\psi^2=\frac{d^2V}{d\psi^2}(v;0)=2\mu_{\rm eff}^2+8Bv^2\ll m_A^2\sim g^2v^2\,.
\end{equation}
The second line in~\eqref{Veff} denotes the thermal correction given by~\cite{Dolan:1973qd,Dine:1992wr,Arnold:1992rz}
\begin{align}\label{thermal_corr}
\Delta V_\mathrm{thermal}({\psi}; T_d) =  \frac{1}{2\pi^2}\sum_i n_{i} T_d^4 J_B(m_i^2(\psi)/T_d^2)\,,
\end{align}
where $J_B(a^2)=\int_0^\infty dp \, p^2 \ln(1-e^{-\sqrt{p^2+a^2}})$
and the sum runs over all species with field-dependent masses $m_{i}$ and $n_i$ degrees of freedom. The parameterization $J_B(a^2)\equiv 2\pi^2K(a)e^{-a}$ from~\cite{Niedermann:2021vgd} makes the Boltzmann suppression for $a\gg1$ explicit, while $K(a) \mathrm{e}^{-a}\simeq - \pi^2/90 + a^2/24 - a^3/(12 \pi)+ \ldots$ in the opposite limit. In the supercooled regime $\lambda\sim g^4$ it is sufficient to include only the  gauge bosons  with masses $m_A\sim gv$ in the sum, and omit Debye corrections known as ring resummation~\cite{Arnold:1992rz}, see~\cite{Gould:2021oba,Kierkla:2023von} for further discussion.
The leading $\psi$-dependent thermal correction for $T_d\gg g\psi$ is
\begin{equation}
  \Delta V_\mathrm{thermal}({\psi}; T_d) \to \frac{1}{24}c_0n_Ag^2T^2\psi^2\,,
\end{equation}
where as before $n_A$ is the number of gauge bosons acquiring mass (see Tab.~\ref{tab:dof}) and $c_0$  another model-dependent constant of order unity ($c_0=4/15$ for $N=3$). In the opposite limit $T_d\ll g\psi$ the thermal correction is exponentially suppressed.

The thermal correction restores the $SU(N)$ symmetry at high temperatures, while a second minimum at non-zero $\psi=v_\psi(T)$ develops and becomes degenerate with the one at $\psi=0$ at the critical temperature $T_c$. For $\mu_{\rm eff}^2\lesssim v^2g^4$, one has\footnote{A more precise value can be easily obtained using the parameterization of $K(a)$ in~\cite{Niedermann:2021vgd} (see also Tab~\ref{table:energy_scales} for a numerical example).}
\begin{equation}\label{Tc}
  T_c\sim gv\,,
\end{equation}
and $v_\psi(T_c)\sim v_\psi(0)=v$. The two minima are separated by a barrier, such that the phase transition is of first order in the regime we consider. The transition temperature $T_d^*$ at which bubbles of the true vacuum are nucleated is significantly below the critical temperature in the supercooled regime, $T_d^*\ll T_c$. In the conformal case $\mu_{\rm eff}=0$, the barrier becomes small but does not vanish (unless the temperature vanishes exactly), leading to potentially very strong supercooling, see e.g.~\cite{Levi:2022bzt,Salvio:2023ynn,Kierkla:2023von}. For non-zero $\mu_{\rm eff}$, the barrier vanishes at a finite temperature, given approximately by the temperature $T_b$ for which $d^2V/d\psi^2(0;T_b)$ changes sign,
\begin{equation}\label{Tb}
  T_b^2 = \frac{12\mu_{\rm eff}^2}{c_0n_Ag^2}= \frac{\gamma}{\pi}g^2v^2\,,
\end{equation}
where we introduced the supercooling parameter~\cite{Niedermann:2021vgd} 
\begin{align}\label{def:gamma}
  \gamma \equiv \frac{12 \pi \mu_{\rm eff}^2}{ c_0 n_A v^2g^4}\,.
\end{align}
The conformal limit of radiative symmetry breaking \`a la Coleman-Weinberg corresponds to $\gamma\to 0$, while supercooling occurs as long as $\gamma\lesssim {\cal O}(1)$. The parameter $\gamma$ can take in principle any value in this range, being arguably technically natural since classical conformal symmetry is restored for $\gamma=0$. Its size controls the amount of supercooling. To determine the temperature at which the phase transition occurs, we can discriminate two regimes. For this it is useful to consider the temperature $T_d^*|_{\rm CW}$ for which tunneling would occur with unit probability in a Hubble volume and time in the Coleman-Weinberg case $\gamma=0$. The first (rather extreme) possibility is that $\gamma$ is so small that $T_b<T_d^*|_{\rm CW}$, and the transition happens essentially at the same time as it would occur for $\gamma=0$. This is the case for $\gamma\ll (T_d^*|_{\rm CW})^2/(gv)^2$. The second case, that we consider in the following, occurs if $\gamma\gtrsim (T_d^*|_{\rm CW})^2/(gv)^2$, such that $T_b>T_d^*|_{\rm CW}$. In this case the $\mu_{\rm eff}^2$ contribution to the effective potential makes the barrier vanish already before the field would have tunneled in the Coleman-Weinberg case. Then, the transition occurs very shortly before the barrier vanishes at temperature $T_b$, i.e.
\begin{equation}
  T_d^* \simeq T_b = \sqrt{\frac{\gamma}{\pi}} gv \sim \sqrt{\gamma}\, T_c\,,
\end{equation}
applicable for $(T_d^*|_{\rm CW})^2/(gv)^2\lesssim \gamma\lesssim 1$, which we assume in the following. 
The parametric dependence for small $g$ can be estimated as~\cite{Witten:1980ez} $T_d^*|_{\rm CW}= v \exp\left( -\mathcal{O}(1)/g^3 \right)$ when using the percolation condition $S^*_3/T_d^* \simeq 250$~(taken from~\cite{Niedermann:2021vgd}, see also~\cite{Salvio:2023ynn}). For $g \lesssim 1$, as assumed in this work, we can thus safely neglect the effect of a non-vanishing $T_d^*|_{\rm CW}$.  

For our purpose, we are interested in the latent heat
\begin{align}\label{def:latent_heat}
\Delta V_* = V({\psi}=v_\psi(T_d^*);T_d^*) -V({\psi}=0;T_d^*)
\end{align}
released in the phase transition. For $\gamma\lesssim 1$ one can estimate the latent heat by the potential energy difference between the two phases in the limit $T=0$, 
\begin{eqnarray}\label{latent_heat}
\Delta V_*&\simeq& V(v;0) -V(0;0)=\frac12Bv^4+\frac14\mu_{\rm eff}^2v^2\nonumber\\
&=&g^4v^4\frac{(3c_1+128\pi c_0\gamma)n_A}{6144\pi^2}
= T_b^4\frac{(3c_1+128\pi c_0\gamma)n_A}{6144\gamma^2}\,,\nonumber\\
\end{eqnarray}
where we used~(\ref{Tb},\ \ref{def:gamma}).

The relevance of having a strongly supercooled phase transition becomes clear when considering the contribution to  $N_\mathrm{eff}$ before and after the phase transition. In terms of the temperature ratio $\xi = T_d / T_\mathrm{vis}$ it is given by (assuming a purely bosonic field content)
\begin{align}\label{def:Neff}
\Delta N_\mathrm{eff} = \frac{4}{7} \left( \frac{11}{4} \right)^{4/3} g_\mathrm{rel,d} \,\xi^4\,,
\end{align}
where $g_\mathrm{rel,d}$ ($g_\mathrm{rel,vis}$) denotes the number of relativistic degrees of freedom in the dark (visible) sector. We further introduce the maximal fraction of vacuum energy stored in the scalar field just before the phase transition, dubbed new early dark energy, as 
\begin{equation}\label{fNEDE}
f_\mathrm{NEDE} = \Delta V_* / \rho_\mathrm{tot}(T_d^*)\,,
\end{equation}
where  $\rho_\mathrm{tot}(T_d^*) \simeq \pi^2 g_\mathrm{rel, vis}^* T_d^{*4}/( 30 \xi_*^4)  + \Delta V_*$ (when neglecting the small contribution of the dark sector radiation plasma before the transition). This allows us to express $\xi_*$, i.e., the relative dark sector temperature right before the phase transition, in terms of $f_\mathrm{NEDE}$ as\footnote{When combining with \eqref{latent_heat}, this reproduces the result in~\cite{Niedermann:2021vgd} for $n_A=3$ and $c_0=1$ when we neglect one-loop corrections (corresponding to the formal limit $c_1 \to 0$). This tree-level approach is valid in the mildly supercooled regime with $1 \gtrsim\gamma \gtrsim 0.03$ (for $N=3$).}
\begin{align}\label{xi_star}
\xi_*^4 \simeq g_\mathrm{rel, vis}^* \frac{\pi^2}{30}  \frac{f_\mathrm{NEDE}}{1-f_\mathrm{NEDE} } \frac{(T_d^*)^4}{\Delta V_*}\,.
\end{align}
Using the result~\eqref{latent_heat} for the latent heat, and $T_b \simeq T_d^*$ in the strongly supercooled regime, we obtain a (small) contribution to $N_\mathrm{eff}$ \textit{before} the phase transition
\begin{align}\label{N_BBN}
\Delta N_\mathrm{BBN} &\simeq \frac{32}{35} \left( \frac{11}{4} \right)^{4/3}    \frac{128\pi^2\gamma^2 g_\mathrm{rel,d}^* g_\mathrm{rel,vis}^*}{ (3c_1+128\pi c_0\gamma)n_A} \frac{f_\mathrm{NEDE}}{1-f_\mathrm{NEDE}}\nonumber\\
& \simeq 0.055 \, \frac{\frac{N^2+N -1}{2N-1} }{11/5}\left(\frac{\gamma}{0.01}\right)^2\frac{52/45}{c_1} \frac{f_\mathrm{NEDE}}{0.08}\,,
\end{align}
where in the second line we used $g_\mathrm{rel, vis}^* = 3.38$, $g^*_\mathrm{rel, d} = 2N^2  + 2N -2$, $n_A=3(2N-1)$ (see Tab.~\ref{tab:dof}) and assumed $\gamma, f_\mathrm{NEDE} \ll 1$. From this, it is clear that if $\gamma$ is sufficiently small, {i.e.}, we are in the strongly supercooled regime, $\Delta N_\mathrm{eff}$ can be made compatible with BBN. As a reference value for $N=3$, we can satisfy the BBN bound $\Delta  N_\mathrm{BBN} \lesssim 0.1$ for $\gamma  \sqrt{f_\mathrm{NEDE}} \lesssim 0.004$ corresponding to $\xi_* \lesssim 0.21 $. For our data fit, we will set $\xi_* = 0.1$, which allows us to satisfy the BBN bound for $N < 15$. As we will argue in detail in the next section, this subdominant dark radiation fluid, through its adiabatic perturbations $\delta_{\rm DR}$, sets the initial conditions for the perturbations, $\delta_\mathrm{NEDE}$ and $\theta_\mathrm{NEDE}$, in the dark sector plasma after the phase transition. We highlight that the precise value of the background parameter $\xi_*$ does not have any impact on the cosmological observables considered here as the corresponding energy density is suppressed by $(\xi_*)^4$.

\subsection{Dark sector reheating} \label{subsec:reheating}

Another central point is the latent heat injection and thermalization of the dark sector after the phase transition.
A first order phase transition leads to the nucleation of vacuum bubbles that separate the false (symmetric) from the true (broken) vacuum. As they expand, more and more space is converted to the true vacuum. Since we are considering strong supercooling, we expect that the latent heat released during this percolation phase is partially stored as kinetic and gradient energy in the bubble walls and partially in coherent scalar field oscillations. Notice that whether supercooling leads to a terminal bubble wall velocity or runaway bubbles demands further analysis, depending on the interaction with the plasma \cite{Bodeker:2009qy,Bodeker:2017cim,Athron:2022mmm,Ellis:2020nnr,Gouttenoire:2021kjv}.  The coupling with the plasma induces sound waves~\cite{Jinno:2020eqg,Jinno:2021ury,Jinno:2022mie,Hindmarsh:2013xza,Hindmarsh:2015qta,Hindmarsh:2016lnk,Hindmarsh:2017gnf} that can reheat the dark sector via other channels (e.g., via heat dissipation in turbulence~\cite{Caprini:2009yp, Cutting:2019zws,Auclair:2022jod}).
As these bubbles start to collide, they break up into smaller fragments and form a field condensate of small-scale anisotropic stress. The corresponding length scale is conventionally denoted in the gravitational wave literature as~\cite{Caprini:2018mtu} $\beta^{-1}$, where $\beta \simeq \dot \Gamma / \Gamma$ quantifies the relative change in the transition rate $\Gamma$.\footnote{To be specific, if the time dependence of $\Gamma$ is approximated as a linear exponential, the parameter $\beta$ corresponds to the inverse duration of the phase transition. For a supercooled transition where the bubble walls quickly approach the speed of light, $\beta^{-1}$, therefore, corresponds to the typical bubble size at the time of collision.} In the supercooled regime, it is given by~\cite{Niedermann:2021vgd} (see also~\cite{Levi:2022bzt}), 
\begin{align}\label{eq:beta}
\beta^{-1} \sim 10^{-2} n_A^{1/3}\, g^2 H(t_*)^{-1} \ll H(t_*)^{-1}\,.
\end{align}
Given a small enough gauge coupling, it is indeed much smaller than cosmological scales set by $H(t_*)^{-1}$. For a more complete description of the gravitational wave phenomenology, see Sec.~\ref{g-ws}.

Depending on the microphysics, there are different possibilities as to how this system can evolve further. The massless gauge bosons offer a channel at one-loop level for the (non-thermal) scalar particles residing in the bubble wall condensate and coherent field oscillations to decay through triangle diagrams into pairs of massless $SU(N\!-\!1)$ gauge bosons, e.g. for $SU(2)$ $ \psi \to A_{1-3} + A_{1-3}$.\footnote{See \cite{Niedermann:2021vgd} for two more possibilities, dubbed scenario A and B. One in which the NEDE boson is stable is expected to inhibit the decay and disintegration of the bubble wall condensate. Another one, which shares similarities with the scenario discussed here, has the bubble wall condensate decay into a stable massive particle that makes a contribution to dark matter.} This is similar to the $h\to \gamma\gamma$ process of the Standard Model. To be precise, we have~\cite{Marciano:2011gm}
\begin{align}
\Gamma^\mathrm{(cm)}_{\psi \to AA} =  g^4 \frac{m_\psi^3}{v^2}  \left( \frac{F}{16 \pi^2}\right)^2 \frac{1}{2\pi}\,,
\end{align}
where $F$ depends on the particle content running in the loop. In our case, where $m_A \gg m_\psi$, we can approximate it as a constant $F=\mathcal{O}(1)$. Setting $N=3$ and employing  \eqref{def:gamma}, \eqref{Tb}, and \eqref{Higgsmass}, we derive    
\begin{align}\label{eq:Gamma}
\Gamma^\mathrm{(cm)}_{\psi \to AA} \simeq    \frac{F^2 g^9  T_d^*}{196608 \sqrt{6} \pi^{15/2}} \frac{(13 + 64\pi \gamma)^{3/2}}{\sqrt{\gamma}} \,. 
\end{align}
At the same time, we have 
\begin{align}\label{eq:H_star}
H_* = \frac{\pi}{\sqrt{90}} \frac{\sqrt{g_\mathrm{rel,vis}^*}}{\sqrt{1 -f_\mathrm{NEDE}}} \frac{T^2_\mathrm{vis*}}{M_\mathrm{pl}}\,.
\end{align}
Combining \eqref{eq:Gamma} and \eqref{eq:H_star}, we obtain the final estimate
\begin{align}\label{eq:Gamma_2}
\frac{\Gamma^\mathrm{(cm)}_{\psi \to AA}}{H_*} = \mathcal{O}(1) \times  \, \frac{g^9 f_\mathrm{NEDE}^{1/4}}{1+z_\ast}10^{24} \,,
\end{align}
where we used $T_\mathrm{vis*} \simeq 2.3 \times  10^{-4} (1+z_\ast) \,\mathrm{eV}$ alongside \eqref{xi_star} and \eqref{latent_heat}, and assumed $f_\mathrm{NEDE}, \gamma \ll1$. Substituting the typical value $f_\mathrm{NEDE}=0.08$, while demanding a post-BBN phase transition with $z_\ast < 10^8$, a sufficient condition for an efficient decay with $\Gamma^\mathrm{(cm)}_{\psi \to AA} \gg H_*$ is $g \gtrsim 0.02$. We note that this calculation assumes that the $\psi$ bosons are at rest at the time of decay, which will only be true right after the bubble nucleation. Also, it neglects bosonic enhancement effects, which might become sizeable. However, we do not expect these effects to significantly tighten the bound on $g$ (after all, it is softened by the ninth root).

While the massless gauge bosons are being populated, they start to thermalize through their self-interactions, $A_{1-3} + A_{1-3} \to A_{1-3} + A_{1-3}$. Their interaction rate is $\Gamma_\mathrm{therm} = \langle\s v\rangle n_h$, where $\langle\s v \rangle$ is the velocity-averaged cross-section. Here, we can employ the much cruder estimate $\Gamma \sim g^4 T_d = g^4 \xi^\mathrm{after}_* T_\mathrm{vis*}$ relying on a (partial) thermalization. With $\xi^\mathrm{after}_* \sim 0.5$, we find
\begin{align}\label{eq:Gamma_3}
\frac{\Gamma_\mathrm{therm}}{H_*} \sim  \frac{g^4}{1+z_\ast}  10^{31}\,,
\end{align}
which, even in the least favourable case with $z_\ast \sim 10^9$ and $g\sim0.01$, ensures an extremely fast thermalization. This also applies to similar number-changing $2\to 3$ or in general $n\to m$ processes involving the massless gauge bosons, that establish complete chemical and kinetic equilibrium with vanishing chemical potential and are parametrically similarly efficient as $2\to2$ processes in a non-Abelian plasma~\cite{Garny:2018grs,Kurkela:2014tea,AbraaoYork:2014hbk}. Once their temperature has reheated above $m_\psi$, the gauge bosons also populate and equilibrate the $\psi$ bosons through the (inverse) decay process $ A_{1-3} +  A_{1-3} \leftrightarrow  \psi $. 

We have therefore established that both the percolation scale $\beta$, as well as the decay and thermalization scales, $\Gamma^\mathrm{(cm)}_{\psi \to AA}$ and $\Gamma_\mathrm{therm}$, are large compared to $H_*$. In the following, we will thus treat the phase transition and subsequent reheating as an instantaneous process on cosmological time scales. As we will see, this simplifies the description of the evolution of both the background fluid and its perturbations.

After the reheating is completed, we can equate the dark sector radiation fluid with the latent heat $\Delta V_*$ (neglecting the small contribution from the preexisting thermal fluid that triggered the phase transition). An analogue calculation as the one described above then yields the contribution to $N_\mathrm{eff}$ \textit{after} the phase transition. To be specific, we find 
\begin{align}\label{eq:xi_after}
\left(\xi^{*,\mathrm{after}}\right)^4=\frac{g_\mathrm{rel, vis}^*}{g_\mathrm{rel, d}^\mathrm{after}}\frac{f_\mathrm{NEDE}}{1-f_\mathrm{NEDE}}\,,
\end{align}
with $g_\mathrm{rel, d}^\mathrm{after}=2[(N-1)^2-1]+1$ (see Tab.~\ref{tab:dof}). Using \eqref{def:Neff}, this translates into

\begin{align}\label{eq:N_NEDE}
\Delta N_\mathrm{NEDE} 
&\simeq  \frac{4}{7} \left( \frac{11}{4} \right)^{4/3}  g_\mathrm{rel, vis}^* \frac{f_\mathrm{NEDE}}{1-f_\mathrm{NEDE}} \nonumber \\
&\simeq 7.4  \frac{f_\mathrm{NEDE}}{1-f_\mathrm{NEDE}}\,.
\end{align}
We note that $\Delta N_\mathrm{NEDE}$ should be identified with  $N_\mathrm{UV}$ as used in the literature on the step model (see for example~\cite{Aloni:2021eaq}).
This shows that we can achieve $\Delta N_\mathrm{NEDE}$ of order unity for $f_\mathrm{NEDE} \simeq 12 \%$ (for $N=3$ this corresponds to $\xi\simeq 0.5$). With the definitions in \eqref{eq:N_NEDE} and \eqref{N_BBN}, we can summarize this \textit{first step} as
\begin{align}\label{eq:step1}
\Delta N_{\rm BBN} \to  \Delta N_{\rm NEDE} =  \frac{5 \left(3c_1 + 128 c_0\pi \gamma\right) n_A}{1024 \pi^2 g_\mathrm{rel,d}^* \gamma^2} \, \Delta N_{\rm BBN} \,.
\end{align}
Therefore, a strongly supercooled phase transition with $\gamma \ll1$ offers an efficient mechanism to heat the dark sector after BBN and introduce a sharp step in $\Delta N_\mathrm{eff}$.

\subsection{The second step} \label{subsec:2ndstep}

\begin{figure}[t]
	\centering
	   \includegraphics[width=0.495\textwidth]{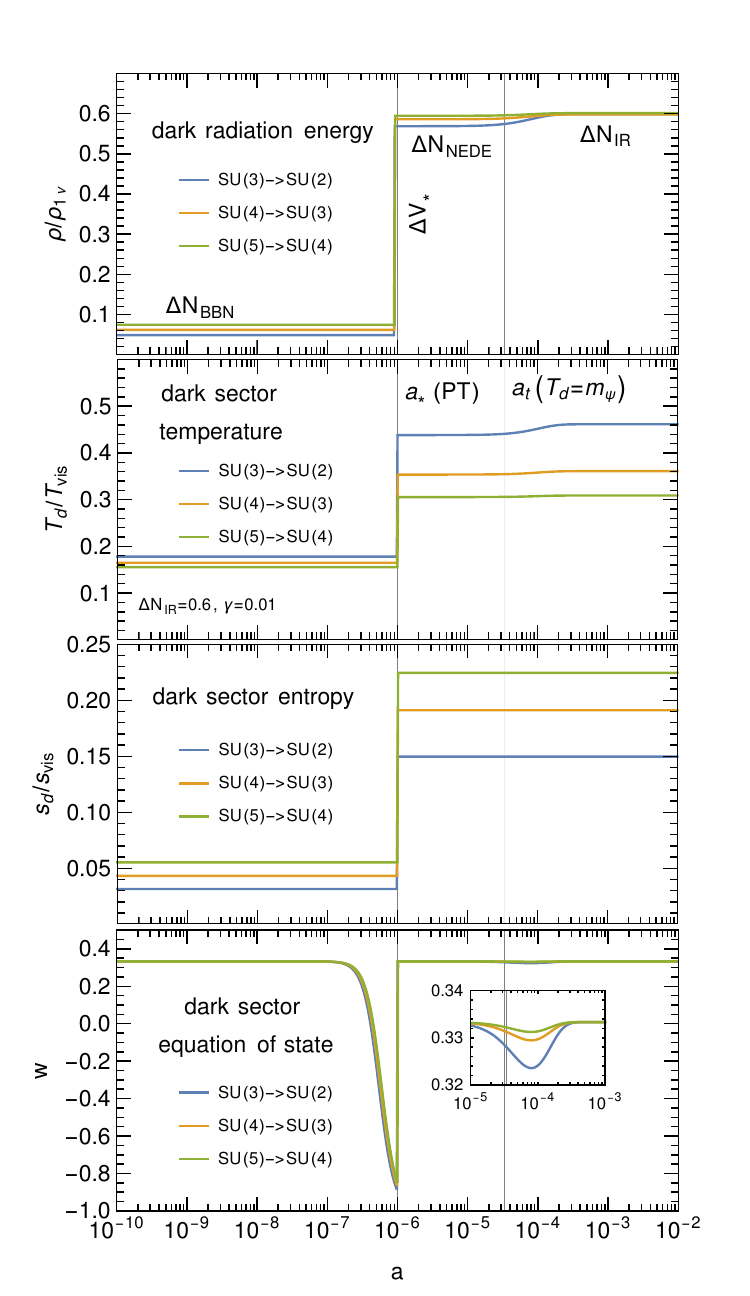}
	\caption{\label{fig:dsentropy}Evolution of $\Delta N_{\rm eff}$, temperature ratio $\xi_d=T_d/T_{\rm vis}$, entropy ratio $s_d/s_{\rm vis}$ and equation of state $w=p_d/\rho_d$ for $SU(N)\to SU(N\!-\!1)$ with $N=3,4,5$. All are discontinuous at the phase transition at $a_*$ (first step), while only the entropy is conserved once the dark Higgs becomes non-relativistic at $a_t$ (second step). The second step disappears in the large-$N$ limit. Note that for a growing number of degrees of freedom $g_d$ (i.e. for $N=3,4,5$), the temperature $T_d$ decreases while $s_d\propto g_dT_d^3$ increases with $g_d$ for given $\Delta N_{\rm IR}\propto g_dT_d^4$. }
\end{figure}

After the phase transition, we have a two-component fluid, with the light dark Higgs bosons and the massless gauge bosons being in thermal and chemical equilibrium. However, when the NEDE bosons become non-relativistic, their production channel becomes Boltzmann suppressed and they decay into the massless gauge bosons, thereby depositing their entropy in the radiation fluid. Microscopically, this happens again through the triangle diagram $\psi \to AA$ and we can use \eqref{eq:Gamma_2} evaluated at redshift $z_t < z_\ast$ (up to an order unity factor accounting for the velocity distribution) to argue that the decay process is efficient and thus this process can be described assuming thermal equilibrium, implying in particular entropy conservation, similarly as for $e^+e^-$ annihilation in the visible sector. As a result, the previous contribution to $N_\mathrm{eff}$ increases further
\begin{align}\label{eq:N_IR}
\Delta N_{\rm NEDE} \to  \Delta N_{\rm IR} = \Delta N_{\rm NEDE} \, (1+r_g)^{1/3} \,,
\end{align}
with $r_g$ the relative change in the number of relativistic degrees of freedom. 
This second step has been studied extensively as a solution to the Hubble tension in the literature~\cite{Aloni:2021eaq,Schoneberg:2022grr} and we review its derivation in the App.~\ref{app:secondstep}. 
For our model, we find $r_g^{-1}=2N (N-2) \geq 6 $, which translates to a small step of less than $5.3 \%$.  We can constrain the redshift $z_t$ of the second step, which is defined implicitly through $T_d(z_t) = m_\psi$. From the definition in \eqref{Higgsmass} and \eqref{latent_heat}, we derive\footnote{As noted before a light Higgs state is one of the characteristics of a supercooled phase transition~\cite{Coleman:1973jx,Witten:1980ez,Gildener:1976ih,Iso:2017uuu}.}
\begin{align}
\frac{m^4_\psi}{\Delta V_*}=\frac{g^4 n_A\left( 3 c_1 + 64 c_0\pi \gamma\right)^2}{24 \pi^2 \left( 3 c_1 + 128 c_0\pi \gamma\right)}\,.
\end{align}
Together with the conditions $(1+z_t)/(1+z_\ast)= m_\psi /T_d^\mathrm{*,after} $ and $\Delta V_* \simeq \frac{\pi^2}{30} g_\mathrm{rel,d} (T_d^\mathrm{*,after})^4 $, this implies
\begin{align}\label{def:z_t}
\frac{1+z_t}{1+z_\ast} 
&\simeq g \left(\frac{c_1 g_\mathrm{rel,d}^\mathrm{after} n_A}{240}\right)^{1/4} \,,
\end{align}
where we assumed $\gamma \ll 1$. 
As an explicit example, in the case of $N=3$, we find that $(1+z_t)/(1+z_\ast) \simeq 0.84\, g \gtrsim 0.02$, where we used our reheating bound $g \gtrsim 0.02$. As a result, the second step is constraint to occur within a few e-folds after the first step. This bound becomes even tighter for $N>3$. However, in this case $r_g\to 0$ and as a consequence the second step  disappears in the large-$N$ limit, see Fig.~\ref{fig:dsentropy}, such that the value of $z_t$ does not play a role in that case. We anticipate that the second step has minor impact in the analysis, being not favoured as a solution to $H_0$ (see \ref{app:resultssecondstep}). This second step is therefore more an extra feature of our model then a necessary point to address the Hubble tension.

The entire two-step sequence, $\Delta N_\mathrm{BBN} \to \Delta N_{\rm NEDE} \to  \Delta N_{\rm IR}$, is depicted in Fig.~\ref{fig:dsenergy} as the purple line.
We stress that the first and second step are physically distinct. While the thermally induced phase transition is first order and thus involves a discontinuous change in the fluid's entropy, the second step is a continuous process for which the entropy remains constant  (see Fig.~\ref{fig:dsentropy}). The remaining evolution at background level is then simply that of a relativistic fluid.

To summarise, the dark sector fluid before the phase transition  ($a < a_*)$ is
\begin{align} \label{eq:rho_before}
\bar{\rho}_\mathrm{DR}(a)  + \Delta V_* \,,
\end{align}
where  $\Delta V_* = \mathrm{const}$ is the latent heat defined in \eqref{def:latent_heat} and $\bar{\rho}_\mathrm{DR} =  g_{\mathrm{rel},d} \frac{\pi^2}{30}  T_d^4 $ denotes the subdominant radiation fluid that triggers the phase transition. The bar notation indicates that we work at the background level. After the bubble percolation and subsequent thermalization have completed, we are left with  a tightly coupled fluid composed of $\psi$ particles and massless gauge bosons with an evolving equation of state parameter, giving rise to ($a > a_*)$
\begin{multline} \label{eq:rho_after}
  \bar{\rho}_{\rm NEDE}(a)  
 = \left[\Delta V_* + \bar{\rho}_\mathrm{DR}(a_*) \right]  \\
 \times \exp{\left[-3 \int_{a*}^a d \ln{\tilde a} \left(1+w(\tilde a)\right) \right]} \,,
\end{multline}
where $w(a)$ is defined in~\ref{eq:w_NEDE}. 
The time evolution for different contributions to the dark sector plasma are depicted in Fig.~\ref{fig:dsenergy} as the orange (latent heat), red (Higgs), and green (massless gauge bosons) lines. \ 

Furthermore, a key assumption in our derivation is that the phase transition completes in much less than a Hubble time. This allow us to describe it as an instantaneous process when implementing our model in a Boltzmann code. In the next section, we discuss how such an abrupt transition affects perturbations. 

\subsection{Perturbations} \label{subsec:perts}

We next turn to perturbations in the dark sector. Before the NEDE phase transition, the subdominant dark radiation fluid carries adiabatic perturbations, while the NEDE vacuum acts like a (homogeneous) cosmological constant. After the phase transition, there are two main sources for perturbations in the radiation fluid. First, there are inhomogeneities related to the nucleation of bubbles of the broken symmetry phase. These perturbations are large and set by the density contrast between the bubble interior and exterior, explicitly $\delta_\mathrm{NEDE} \sim 1$ (as we can neglect the radiation plasma before the transition). However, they are only present on small scales $1/\beta$ set by the size of vacuum bubbles before they collide. In particular, for a phase transition that occurs much before recombination and $g$ sufficiently small, these modes cannot be probed neither in the CMB nor in LSS data.\footnote{For concreteness, a sufficient condition for the absence of any direct bubble signature in CMB data is~\cite{Niedermann:2020dwg} $H_*/\beta \ll 0.01 (z_*/5000)$ (see also~\cite{Elor:2023xbz} for a recent discussion). According to \eqref{eq:beta}, this condition only requires $g\lesssim {\cal O}(1)$ even for a $z_\ast$ as low as $10^4$.}  Second, there are perturbations seeded by adiabatic perturbations of the (subdominant) radiation fluid that trigger the thermal phase transition. Let us stress that these perturbations are not a consequence of the stochastic character of the phase transition; instead, they are created because different spatial regions are slightly ``ahead'' or ``behind'' in time. In a coarse-grained approach valid on large scales, i.e., for spatial momenta $k\equiv|\mathbf{k}| \ll \beta^{-1} $, they can be derived by matching the cosmological perturbation theory across a surface of constant $\delta_\mathrm{DR}(\eta,\mathbf{k}) \equiv [\rho_\mathrm{DR}(\eta,\mathbf{k}) - \bar{\rho}_\mathrm{DR}(\eta)]/\bar{\rho}_\mathrm{DR}(\eta)$.\footnote{This instantaneous matching approach has also been shown to work in an inflationary context where the inflation sets the initial perturbations in the cosmic fluid after reheating~\cite{Deruelle:1995kd}. The main conceptual difference here is that we must also track subhorizon modes as our phase transition occurs at a later time when observable scales have already started to enter the horizon.}
The details of this matching procedure are described in~\cite{Israel:1966rt,Niedermann:2020dwg,Niedermann:2021vgd}, giving rise to the initial conditions for the NEDE density contrast $\d_\mathrm{NEDE} \equiv \delta \rho_\mathrm{NEDE} / \bar{\rho}_\mathrm{NEDE}$ and velocity divergence $\theta_\mathrm{NEDE}$ valid in synchronous gauge (following the definitions in Ma and Bertschinger~\cite{Ma:1995ey})
\begin{subequations}
\label{eq:matching}
\bea
\delta_{\rm NEDE}^\mathrm{(+)} &=& -3 \left(1+w^\mathrm{(+)}\right) \mathcal{H}_\ast \frac{\delta q_\ast}{\bar{q}'_\ast} \vs 
&=& \frac{3}{4}\left(1+w^\mathrm{(+)}\right)\delta_{\rm DR}^\ast   \,, \\
\theta_{\rm NEDE}^\mathrm{(+)} &=& k^2\frac{\delta q_\ast}{\bar{q}'_\ast} =-\frac{k^2}{4\mathcal{H}_\ast}\delta_{\rm DR}^\ast  \,, 
\eea
\end{subequations}
where $\mathcal{H} = a H$, primes denote derivatives w.r.t.\ conformal time $\eta$, and $(+)$ indicates evaluation right \textit{after} the phase transition. The variable $q(\eta,\mathbf{x}) \equiv \bar{q}(\eta) + \delta q(\eta,\mathbf{x})$ defines a general transition surface leading to spatial trigger time variations $\delta \eta(\mathbf{x}) = - \delta q_\ast / \bar{q}'_\ast$. In accordance with the discussion in the previous sections, in the case of Hot NEDE, a surface of constant temperature corresponds to a surface of constant tunneling probability. We thus identify $ q_*(\mathbf{x})$ with the dark sector temperature $ T_d(\eta_*,\mathbf{x})$. This allows us to relate  
\bea\label{eq:delta_q}
\frac{\delta q_\ast}{\bar{q}'_\ast} = -\frac{1}{\mathcal{H}_\ast} \frac{\delta T_d^*}{T_d^*} = -\frac{1}{4\mathcal{H}_\ast} \delta_{\rm DR}^\ast\,,
\eea
which was used in \eqref{eq:matching}.
In a scenario like ours where all the latent heat is converted to radiation, we have $w(a \geq a_*)\simeq 1/3$. As a result, $\delta^\mathrm{(+)}_\mathrm{NEDE} \simeq \delta^*_\mathrm{DR}$ and $\theta^\mathrm{(+)}_\mathrm{NEDE} \simeq -k^2  \delta_\mathrm{DR}^* /(4 \mathcal{H}_*)$. It is the latter condition that leads to the matching's distinct imprint on subhorizon scales. We stress that this makes the Hot NEDE phenomenology differ from the SIDR case for modes that have entered the horizon before the phase transition.

For their subsequent time evolution, we use the  continuity and Euler equation in synchronous gauge~\cite{Ma:1995ey,Hu:1998kj,Blas:2011rf},
\begin{subequations} \label{eq:fluid_equations}
\begin{align}
{\d}^\prime_{\rm NEDE} &= -\left( 1+ w \right)\left(\theta_{\rm NEDE} + \frac{h^\prime}{2}\right) \nonumber\\ &\quad\quad\quad\quad\quad\quad - 3 \mathcal{H} \left(c_s^2 - w \right) \delta_\mathrm{NEDE}\,,\label{eq:pertd}\\
{\theta}^\prime_{\rm NEDE} &=  \frac{k^2 c_s^2}{1 + w^2} \d_{\rm NEDE} - \mathcal{H} \left(1-3 c_s^2  \right) \theta_\mathrm{NEDE}\,, 
\end{align}
\end{subequations}
where $h$ is the spatial trace of the metric perturbation, and the sound speed $c_s(a)$ is defined in \eqref{eq:cs_NEDE}.
The self-interaction required to suppress higher moments and ensure fluid-like behaviour is naturally realized within the dark sector model by the residual gauge interactions of the $SU(N\!-\!1)$ group after the phase transition as well as the $\psi\leftrightarrow AA$ process discussed above.

\section{Cosmological parameter extraction}\label{data}

\subsection{Implementation and data sets}\label{sec:data}

We implemented the phenomenological model described in the previous sections into \texttt{TriggerCLASS}\footnote{\href{https://github.com/NEDE-Cosmo/TriggerCLASS}{https://github.com/NEDE-Cosmo/TriggerCLASS}}, which builds on the Cosmic Linear Anisotropy Solving System (\texttt{CLASS})~\cite{Blas:2011rf}. The background radiation fluid that triggers the NEDE phase transitions is initialised with a fiducial dark sector temperature $\xi=0.1$ to ensure the suppression of $\Delta N_\mathrm{BBN}$. As previously mentioned, this initial dark sector is very subdominant and its temperature has no impact for observations. Nevertheless its perturbations play an important role for triggering the perturbations of the dark sector fluid after the phase transition. The first step in $\Delta N_\mathrm{eff}$ is characterized by its redshift $z_\ast$ and latent heat fraction $f_\mathrm{NEDE}$ [fixing $\Delta N_\mathrm{NEDE}$ via \eqref{eq:N_NEDE}]. Similarly, the second step is determined by its redshift $z_t$ and size $r_g$ [fixing $\Delta N_\mathrm{IR}$ through \eqref{eq:N_IR}]. At the level of perturbations, the trigger fluid is initialised as a tightly-coupled radiation plasma with adiabatic perturbations $\delta_\mathrm{DR}$ and $\theta_\mathrm{DR}$, which are evolved using the standard fluid perturbation equations with $w = c_s^2 = 1/3$. The matching across the transition surface uses \eqref{eq:matching} to initialise the NEDE perturbations $\delta_\mathrm{NEDE}$ and $\theta_\mathrm{NEDE}$, which are subsequently integrated using the system \eqref{eq:fluid_equations} (with more details provided in Appendix~\ref{app:secondstep}). In total, this extends $\Lambda$CDM by four parameters: $f_\mathrm{NEDE}$, $z_\ast$, $r_g$ and $z_t$. However, for reasons that will become clear later, we fix $r_g=0$ (which renders $z_t$ unconstrained) and $z_\ast = 10^6$ in most of our analysis, which leaves us with $f_\mathrm{NEDE}$ (or $\Delta N_\mathrm{NEDE}$ equivalently) as the only parameter. We will compare our model with the SIDR model, which extends $\Lambda$CDM with $\Delta N_\mathrm{eff}$ as one parameter. The strongly coupled nature of the DR component is as usual implemented by using a fluid description with vanishing shear.

For the Markov chain Monte Carlo (MCMC) analysis, we use the Python code \texttt{MontePython}~\cite{Audren:2012wb,Brinckmann:2018cvx} interfaced with \texttt{TriggerCLASS} and run the Metropolis Hastings algorithm. We vary all six $\Lambda$CDM parameters, {\it i.e.}\ the dimensionless dark matter and baryon densities $\omega_\mathrm{cdm}$ and $\omega_\mathrm{b}$, the dimensionless Hubble parameter $h$, the reionization depth $\tau_\mathrm{reio}$, as well as the spectral tilt and amplitude $n_s$ and $A_s$ with standard prior choices. We further impose the prior $0 < f_\mathrm{NEDE} <0.3$, and if applicable also $2 < \log_{10}(z_\ast) <  6$, $0< \log_{10}(z_\ast/z_t)<3$, and $-2 < \log10(r_g) < 3$. We consider chains to be converged when the Gelman-Rubin parameter satisfies $R-1 < 0.05$. All neutrinos are effectively treated as massless by setting $N_\mathrm{eff}= 3.044$. Unless stated otherwise, we have used the \texttt{CLASS} interpolation table, computed with \texttt{PArthENoPE v1.0}~\cite{Pisanti:2007hk}, to infer the primordial helium abundance $Y_p$ as a function of $N^\mathrm{BBN}_\mathrm{eff}\equiv3.044+\Delta N_{\rm BBN}$ and $\omega_b h^2$.

\begin{figure}[t]
	\centering
	\includegraphics[width = 0.45\textwidth]{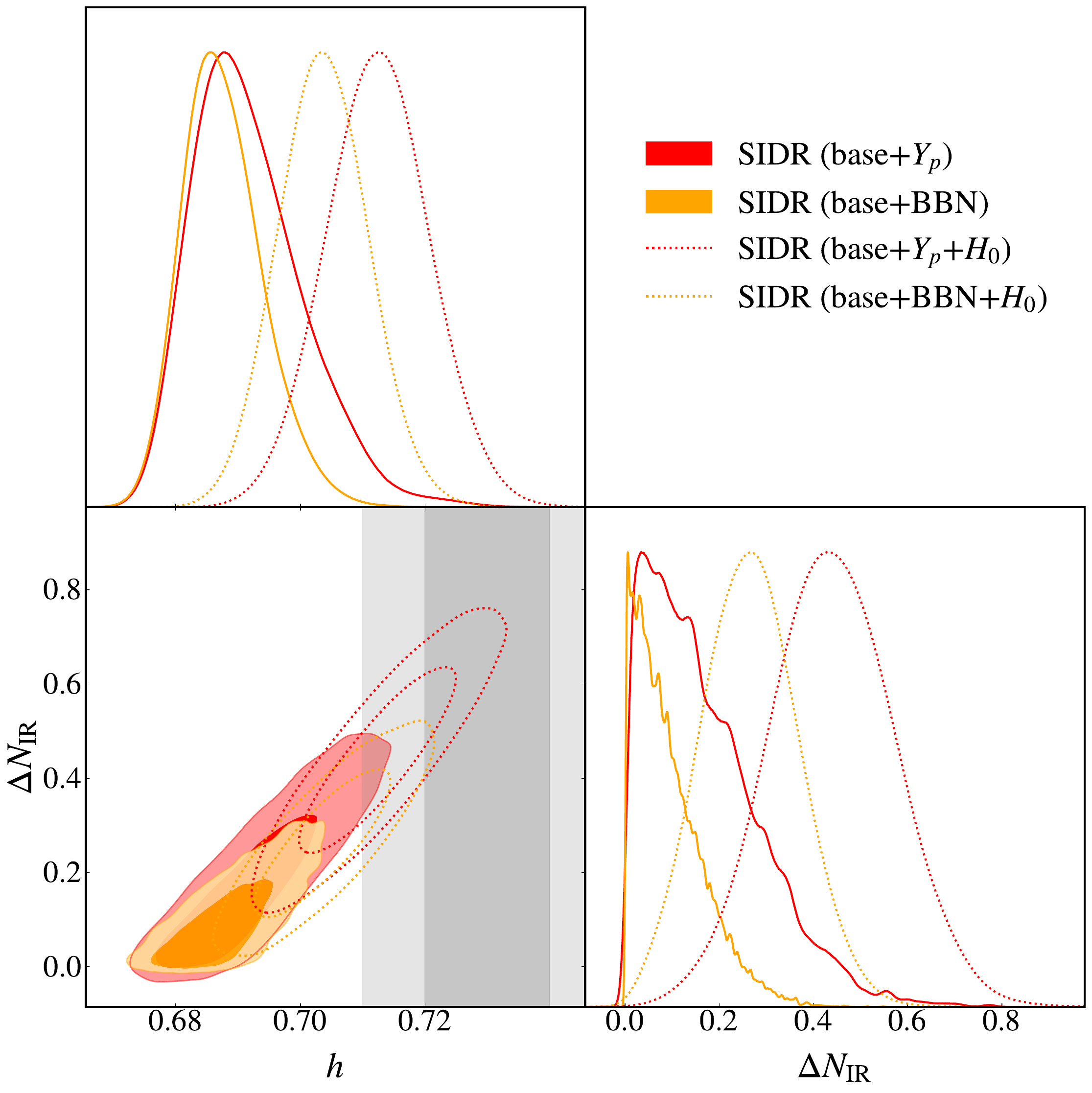}
	\caption{Impact of BBN constraint on a simple SIDR model with constant $\Delta N_{\rm eff}$. Orange contours show $68\%$ C.L. and $95\%$ C.L. marginalized posteriors when including the BBN prior on $\Delta N_{\rm eff}$ and using the corresponding helium abundance predicted by BBN. The red contours show for comparison the case when ignoring BBN constraints and setting the helium fraction  to the value found for $\Lambda$CDM ($Y_p = 0.2454$) by hand, as done in previous analyses. Filled contours correspond to the case without SH$0$ES, and open dotted contours illustrate the impact of including SH$0$ES. We observe that BBN disqualifies SIDR models without post-BBN heating mechanism for addressing the $H_0$ tension.
 }
	\label{fig:resultsSIDR}
\end{figure}

\begin{figure}[t]
	\centering
	\includegraphics[width = 0.5\textwidth]{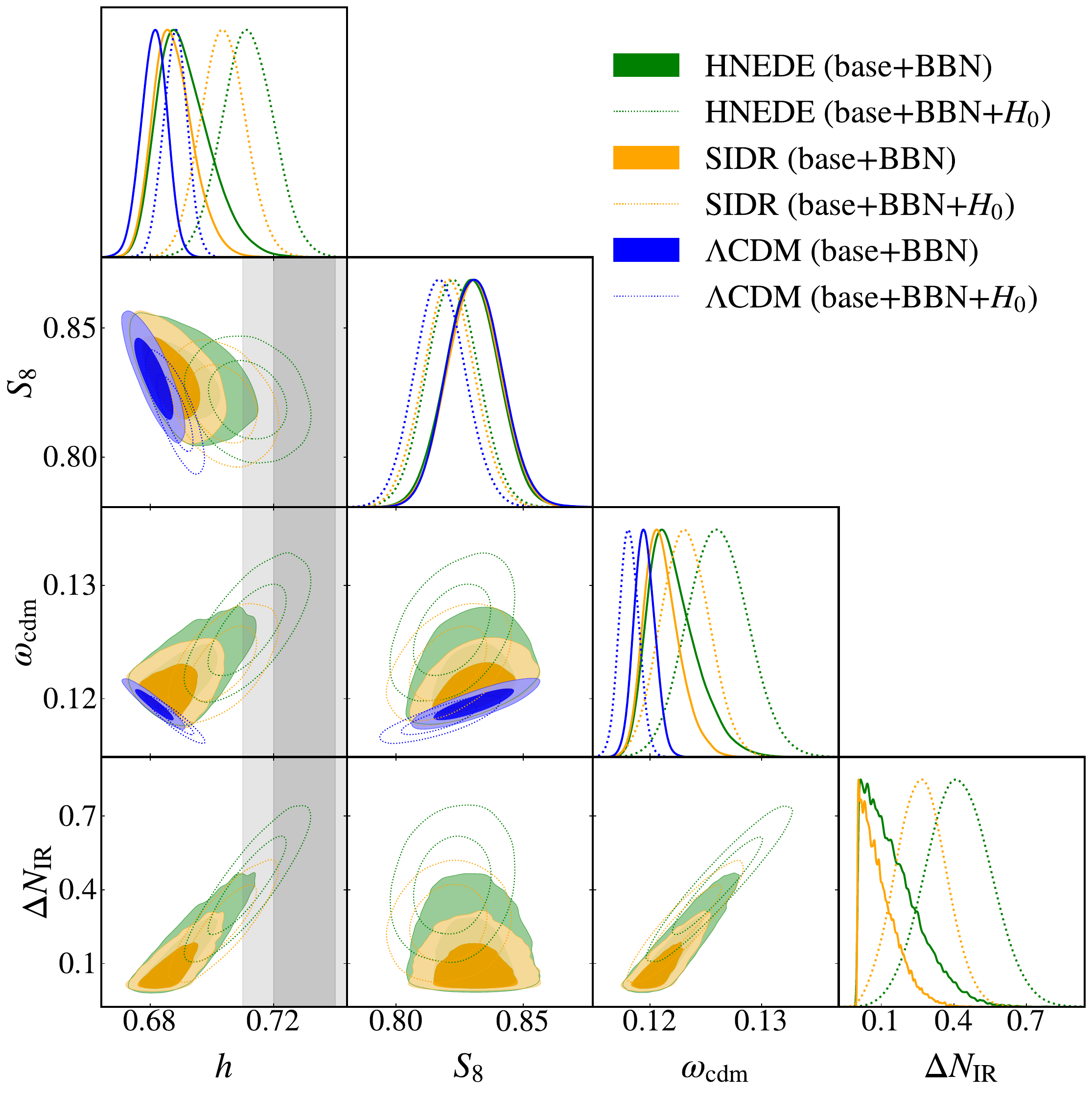}
	\caption{Comparison between the $\Lambda$CDM (blue), SIDR (orange), and Hot NEDE (green) $68\%$ C.L. and $95\%$ C.L. marginalized posteriors, with BBN information included in all cases, and without (filled) and with (open dotted) including SH$0$ES. Only Hot NEDE is marginally compatible with the SH$0$ES value of $h$ (gray vertical band). }
	\label{fig:resultsHNEDEcomparaSIDE}
\end{figure}

We use the following data sets for our analysis:\footnote{Here we closely follow the data sets considered in~\cite{Aloni:2021eaq}, leaving the inclusion of information about the full-shape of the matter power spectrum, redshift space distortions, and additional ground-based CMB data for future work.}
\begin{itemize}
\item \bf Planck 2018\rm: The combined {\it Planck 2018} lensing, high-$\ell$ TT+TE+EE and low-$\ell$ TT+EE CMB anisotropy measurements~\cite{Planck:2018vyg,Planck:2018lbu}. 

\item \bf BAO\rm: We include three BAO measurements from 6dFGRS at $z=0.106$ \cite{Beutler:2011hx}, SDSS at $z=0.15$ \cite{Ross:2014qpa} and also from SDSS-III DR12 at $z=0.38$, $0.51$, $0.61$~\cite{BOSS:2016wmc}.

\item \bf Pantheon\rm: The `Pantheon Sample' luminosity distances of SNe Ia covering the redshift range $0.01 < z < 2.3$ \cite{Pan-STARRS1:2017jku}.

\item \bf SH0ES\rm: To assess the ability of the model to address the $H_0$ tension, in addition to our main analysis, we also use the SH0ES results from the Cepheid–SN Ia sample at $z<0.01$ to put priors on the Hubble parameter $h = 0.7304 \pm 0.0104$ \cite{Riess:2021jrx}.

\item \bf BBN\rm: We include the BBN information in the form of a prior on $\Delta N_{\nu}^{\rm BBN} = 2.889 \pm 0.229$ obtained from helium and deuterium abundance measurements~\cite{Yeh:2022heq}, noticing that this result is CMB-independent, i.e., it does not include the CMB measurements of the baryon abundance or use the CMB constraint on $\Delta N_\mathrm{eff}$.\footnote{We use the relation $\Delta N_{\rm eff} = (1.0147) \Delta N^{\rm BBN}_\nu$.}

\end{itemize}

We refer to Planck 2018, BAO and Pantheon as our ``base'' data set. When quantifying the Hubble tension, we will use the \textit{difference of the maximum a posteriori} measure
\begin{align}
Q_\mathrm{DMAP}(H_0) = \sqrt{ \chi^2_\mathrm{w\backslash\, H_0} -\chi^2_\mathrm{w\backslash o\, H_0} }\,. 
\end{align}
Following~\cite{Raveri:2018wln}, we compute it as the root square of the difference between the bestfit $\chi^2$ values with and without the SH0ES prior on $H_0$ included.  We further employ the Akaike Information Criterion~\cite{Liddle:2007fy} $\Delta \mathrm{AIC} = \Delta \chi^2 + 2 \times (\# \text{added parameters})  $ for model comparison.

\subsection{Results}\label{sec:results}

Before presenting results for the Hot NEDE model, we first discuss the impact of taking into account BBN constraints on $\Delta N_\mathrm{eff}$  and $Y_p$ within SIDR. The purpose of this exercise is to quantify in how far models that do {\it not} feature a post-BBN heating mechanism are disfavoured when it comes to the $H_0$ tension.

\subsubsection{SIDR}

Different variants of the stepped and non-stepped SIDR model have been investigated rather extensively in the literature as a possible solution to the Hubble tension~\cite{Schoneberg:2021qvd,Aloni:2021eaq,Joseph:2022jsf,Schoneberg:2022grr,Allali:2023zbi,Schoneberg:2023rnx,khalife2024review}. Therefore, it is not our ambition here to repeat these analyses or add to the controversy about their effectiveness as solutions to the Hubble tension when BBN constraints are absent (for extensive recent studies, see ~\cite{Allali:2023zbi,Schoneberg:2023rnx}). Instead, we will focus on a rather basic SIDR model without dark matter interactions and mass threshold and with constant $\Delta N_\mathrm{eff}$ throughout the relevant BBN and recombination epochs. We add a BBN constraint on $\Delta N_\mathrm{eff}$ to the base data sets when analysing the SIDR model. Our corresponding results are summarized in Fig.~\ref{fig:resultsSIDR}. We see that including the BBN constraint on $\Delta N_\mathrm{eff}$ (orange contour) makes the SIDR model {\it incompatible} with the SH0ES value of $H_0$ (gray vertical band). This corresponds to a residual DMAP tension of $3.9 \sigma$ (only slightly improving on the $\Lambda$CDM tension of $4.3 \sigma$). Correspondingly the fit improvement over $\Lambda$CDM remains relatively small with $\Delta \chi^2 = -4$ when the analysis includes the (incompatible) $H_0$ prior. These findings clearly remove the simple SIDR model from the space of interesting proposals to resolve the Hubble tension. Let us stress that similar conclusions were reached for the stepped SIDR model when BBN constraints were included~\cite{Schoneberg:2022grr}. Nevertheless,  most of the previous studies on the (stepped) SIDR solutions find smaller values for the residual tension as they neither infer $Y_p$ from BBN nor impose constraints on $\Delta N_\mathrm{eff}$. However, we take the point of view that a viable model should be able to describe both the BBN and CMB epochs, since they interfere in multiple ways. This motivates models featuring a post-BBN heating mechanism such as the Hot NEDE scenario investigated in this work, to which we turn next.

\begin{table*}[ht]
\renewcommand{\arraystretch}{1.3}
	\centering
		\begin{tabular}{ | l || c | c | c | c |  c || c | c | c | c | c ||c|}
			\hline
			\multirow{2}{*}{} & 
			\multicolumn{5}{|c||}{ { Base} + BBN} &
			\multicolumn{5}{|c||}{ + $H_0$} & 
            \multirow{2}{*}{{\scriptsize$\sqrt{Q_{\rm DMAP}^{H_0}}$}}\\ 
			\cline{2-11}
			&$H_0$&$\Delta N_{\rm IR}$&$\chi^2$&$\Delta \chi^2$&$\Delta$AIC&$H_0$&$\Delta N_{\rm IR}$&$\chi^2$&$\Delta \chi^2$&$\Delta$AIC& \\ \hline
			
			$\Lambda$CDM  &$68.13\pm 0.42$& - & 3810.5 & - & - &$68.81\pm 0.39$& -& 3829.7   & - & -&4.3$\sigma$
			\\ \hline 
			SIDR  &$68.77^{+0.52}_{-0.73}$ &$0.094^{+0.024}_{-0.093}$&3810.5\footnote{Here we take the $\Lambda$CDM value as bestfit, since we were not able to find the bestfit for SIDR with the base datasets.} & 0.0 & 2.0 &$70.37\pm 0.72$& $0.27\pm 0.10$ & 3825.7 & -4.0 & -2.0 &3.9$\sigma$
			\\ \hline 

   			Hot NEDE &$69.13^{+0.62}_{-1.0}$&$0.151^{+0.041}_{-0.15}$& 3810.4  & -0.1 & 1.9 & $71.17\pm 0.83$ &$0.42\pm 0.13$& 3818.3 &-11.4 & -9.4 &2.8$\sigma$ \\

			\hline

		\end{tabular}
	\caption{Summary of results for our combined analyses (with and without a SH0ES prior on $H_0$). We compare $\Lambda$CDM, SIDR, and Hot NEDE (with negligible second step, i.e., $r_g=0$). If applicable, we present the mean values with $\pm 1 \sigma$ error. As a result of including the BBN constraints, the SIDR model is hardly an improvement over $\Lambda$CDM. 
	}
	\label{tab:resultstension}
\end{table*}

\begin{table*}[ht]
\renewcommand{\arraystretch}{1.3}
	\centering
		\begin{tabular}{ | l || c | c ||   c ||  c | c | c| }
			\hline
			\multirow{2}{*}{} & 
			\multicolumn{2}{|c||}{ $\Lambda{\rm CDM}$} &
			\multicolumn{1}{|c||}{SIDR}  &
			\multicolumn{3}{|c|}{Hot NEDE}\\
			\cline{2-7}
			&Base+BBN&+$H_0$ &Base+BBN+$H_0$& Base+BBN &+$H_0$ &+$H_0$, $r_g$, $z_{t}$  \\ \hline
			
{\it Planck} High $\ell$ TT-TE-EE & 2350.4 & 2350.6 & 2354.6 & 2351.7 & 2355.1 & 2353.0\\
{\it Planck} Low $\ell$ EE & 396.0 & 398.5 & 398.1 & 395.7 & 395.8 & 398.3\\
{\it Planck} Low $\ell$ TT & 23.7 & 23.0 & 22.3 & 22.7 & 22.2 & 21.5\\
{\it Planck} lensing & 8.8 & 8.6 & 8.9 & 8.9 & 9.2 & 9.4\\
BAO    & 5.2 & 6.1 & 6.0 & 5.2 & 6.2 & 7.2\\
SH0ES    & - & 15.6 & 7.2 & - & 3.7 & 1.5 \\
Pantheon & 1025.9 & 1026.6 & 1025.77 & 1025.7 & 1025.6 & 1026.1\\
BBN    & 0.2 & 0.2 &  2.5 & 0.2 & 0.2 &0.2 \\
\hline	
Total  & 3810.5 & 3829.7 & 3825.7 & 3810.4 & 3818.3 & 3817.5 \\

			\hline	
			
		\end{tabular}
	\caption{Detailed comparison of the bestfit $\chi^2$ for different datasets and models, including BBN information in all cases. For Hot NEDE, we also include the case when allowing for the ``second step`` in the last column, which performs similarly than our fiducial analysis with $r_g=0$. Note that we do not include the case with free $z_\ast$, since it is very similar to the fiducial scenario with $z_\ast=10^6$. The SIDR model without $H_0$ prior is also not shown, see Tab.\,\ref{tab:resultstension}. 
		\label{tab:chi2formodels} }
\end{table*}

\subsubsection{Hot NEDE}

In the Hot NEDE model  $N_\mathrm{BBN} \simeq 3.044$ during BBN and thus including the BBN constraint does not lead to any penalty. Thus, while both the Hot NEDE set-up studied in this work as well as SIDR lead to a similar $\Delta N_{\rm eff}$ around recombination, the BBN constraint hardly affects the posteriors within Hot NEDE, and (as expected) merely creates a small, data-insensitive offset $\Delta \chi^2_\mathrm{BBN} = 0.23$.

We provide a comparison between the $\Lambda$CDM, Hot NEDE and SIDR posteriors in \ref{fig:resultsHNEDEcomparaSIDE}, with the dashed contours representing the addition of the $H_0$ prior. We find that Hot NEDE provides the larger values of $H_0$, {\it when including BBN information}.
As the main result, described in Tables \ref{tab:resultstension} and \ref{tab:chi2formodels}, we find that the DMAP tension is reduced to $2.8 \sigma$ (which should be compared to $3.9 \sigma$ for the SIDR model). In the absence of the SH0ES prior on $H_0$, we obtain $H_0 = 69.13^{+0.62}_{-1.0} \, \mathrm{km}\, \mathrm{sec}^{-1} \mathrm{Mpc}^{-1}$ ($68\%$ C.L.), which translates into a Gaussian tension of $3.2 \sigma$. This is slightly higher than the DMAP tension of $2.8 \sigma$, which we attribute to the non-Gaussian shape of the posterior.
On the other hand, if we include the SH0ES prior, we find $\Delta N_\mathrm{IR}= 0.42 \pm 0.13$ ($68\%$ C.L.), which corresponds to a larger than $3 \sigma$ preference for Hot NEDE over $\Lambda$CDM. While the model in this simple form is arguably not a full resolution of the Hubble tension, our point here is that the model remains in the space of \textit{possible} solutions, whereas DR models susceptible to the BBN constraint are not. Here we consider a simple and minimal model and we leave for future work to investigate extended versions that, for example, feature interactions between the dark radiation and dark matter components. 

A further interesting feature that can be observed in Fig.\,\ref{fig:resultsHNEDEcomparaSIDE} is that for the Hot NEDE model and the base+BBN dataset, $S_8$ slightly decreases when increasing $h$. We checked that this can be attributed to a decrease in the matter density parameter $\Omega_{\rm m}=(\omega_{\rm cdm}+\omega_b)/h^2$ entering $S_8=\sigma_8\sqrt{\Omega_{\rm m}/0.3}$, while the clustering amplitude on scales of $8\,h/$Mpc, $\sigma_8$, stays approximately constant or even slightly increases with $h$. The decrease of $\Omega_{\rm m}$ is a consequence of an over-compensation of the increase of $\omega_{\rm cdm}$ due to the factor $1/h^2$. We note that these features could be interesting to investigate further in view of the $S_8$ tension reported by weak lensing shear measurements, see e.g.~\cite{Garcia-Garcia:2024gzy} for a recent analysis.

\begin{figure}[t]
	\centering
	\includegraphics[width = 0.5\textwidth]{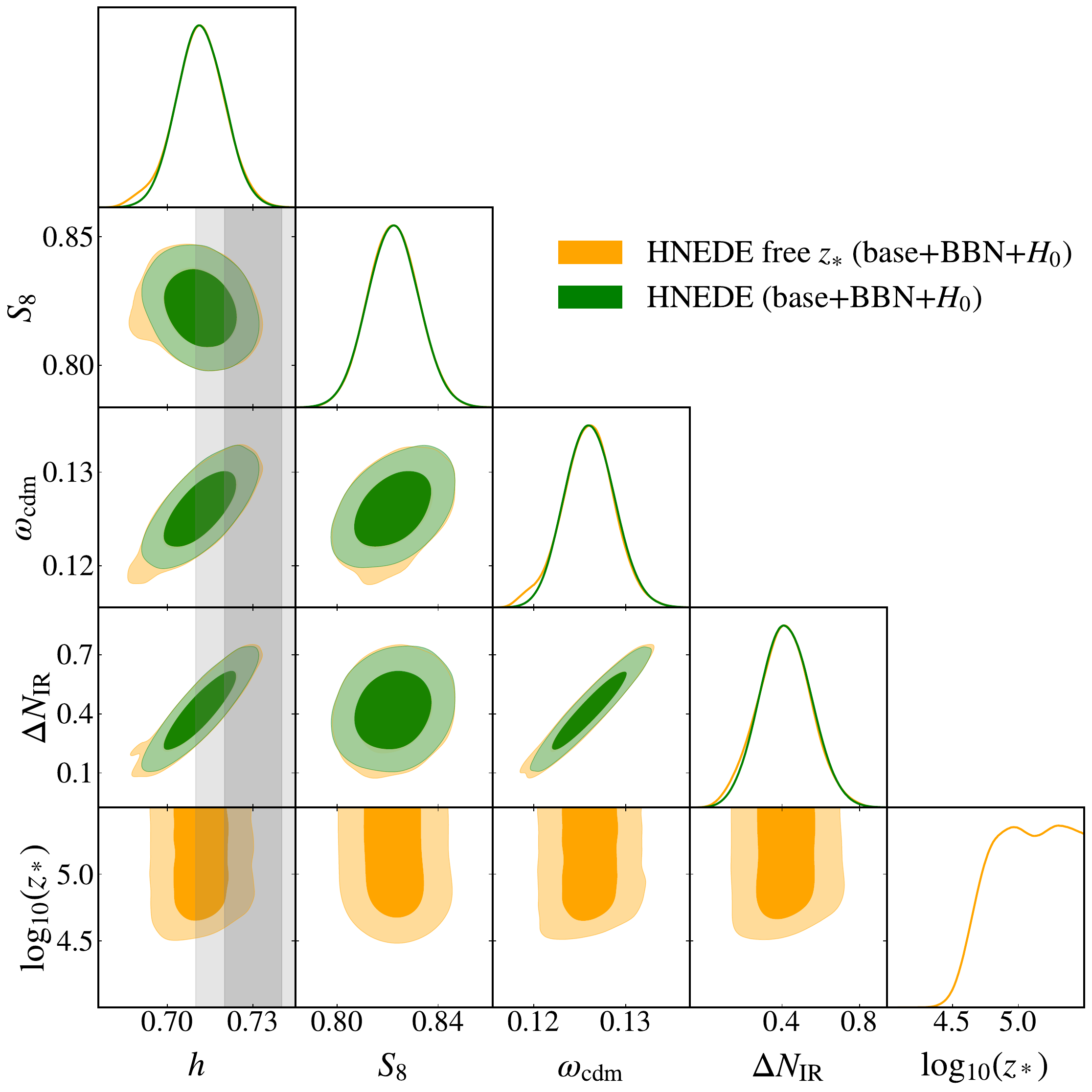}
	\caption{
 Posteriors when including the redshift of the dark sector phase transition $z_\ast$ as a free parameter, compared to the fiducial case with fixed $z_\ast=10^6$. Including $z_\ast$ does not affect the posteriors of the other model parameters, and provides us with a lower bound $z_\ast > 4.4 \times 10^4 $ ($95\%$ C.L.). }
	\label{fig:resultsHNEDE_zast}
\end{figure}

We note that, when choosing the redshift of the phase transition $z_\ast\gtrsim 10^5$ such that modes affecting the CMB were still outside the horizon at $z_\ast$, we  expect that the Hot NEDE model yields results that are comparable to previous analyses of SIDR models that had ignored BBN constraints. We find that this is indeed true for our fiducial choice $z_\ast=10^6$. Comparing the green contours in Fig.~\ref{fig:resultsHNEDEcomparaSIDE} with the red ones in Fig.~\ref{fig:resultsSIDR}, we see that there is almost no discernible difference between both models {\it when the BBN constraint on SIDR is removed}.\footnote{This also serves as a non-trivial cross-check of how we initialize perturbations in the NEDE fluid on super-horizon scales.} 

Similarly, data is also not able to meaningfully constrain the second step (green contour in Fig.~\ref{fig:resultsHNEDE}).  This is broadly in agreement with the findings in~\cite{Allali:2023zbi}, although it is possible that in a slightly more sophisticated scenario with dark matter interactions the step feature can be more clearly constrained. As this is not the focus of the current analysis, we leave this question for future explorations, and, for simplicity, set $r_g=0$ (which also renders $z_t$ unconstrained). For a broader discussion on the second step, see App.~\ref{app:resultssecondstep}.

Moreover, as we demonstrate in Fig.~\ref{fig:resultsHNEDE_zast}, data  provides us with a lower bound $\log_{10}(z_\ast) > 4.64$ ($95\%$ C.L.) (orange contour). This is an important result, as it yields a wide window (roughly five orders of magnitude in redshift) between BBN and $z=4.4 \times 10^4$ during which the phase transition can have occurred. We note that this result also highlights the relevance of the trigger physics. For comparison, in the case of Cold NEDE, adiabatic perturbations in a slowly rolling scalar field trigger the phase transition; instead, here, this role is played by perturbations in the dark radiation fluid. While this difference in microphysics would not matter on superhorizon scales, it does make a difference on subhorizon scales. Although we do not provide a quantitative exploration, the qualitative explanation is as follows: The thermal trigger enhances the initial conditions for $\delta_\mathrm{NEDE}$ and $\theta_\mathrm{NEDE}$ on scales $k > \mathcal{H}_* $ via \eqref{eq:matching} [together with \eqref{eq:delta_q}]. As a result, these modes will carry rather strong dark acoustic oscillations, which would be incompatible with the CMB if the phase transition would occur at $z_\ast\lesssim 10^5$. This is illustrated in  Fig.~\ref{fig:pk} below, where we compare the CMB temperature power spectrum for Hot NEDE cosmologies for different values of $z_\ast$.  However, this conclusion is avoided if the phase transition occurs early enough such that the CMB modes are still frozen on superhorizon scales at $z_\ast$. This is different for Cold NEDE. Here, the dark acoustic oscillations are moderated by the amount of slow-rolling of the trigger field at the time of the phase transition~\cite{Niedermann:2020dwg}. In Sec.~\ref{sec:smallscales}, we argue that the enhanced dark acoustic oscillations imprinted on somewhat smaller scales than those probed by the CMB are actually an interesting signature of Hot NEDE that can be searched for in LSS data, and that further sets it apart from SIDR models.

As a result, we are left with a one-parameter extension of $\Lambda$CDM that performs similarly as analyses of the SIDR model when disregarding BBN. When including BBN constraints, Hot NEDE remains largely unaffected, while SIDR is disfavoured. Thus the heating by the first order phase transition provides an efficient mechanism to make models addressing the Hubble tension within an SIDR-like set-up consistent with BBN.

\section{Phenomenology}\label{sec:pheno}

A vacuum phase transition between BBN and recombination leads to unique signatures. If the phase transition happens early enough with $z_\ast > 10^6$ there is the prospect of detecting a stochastic background of low-frequency gravitational waves. On the other hand, if it does occur sufficiently late with $z_\ast < 10^6$, it will imprint itself in the CMB and in the matter power spectrum on small scales. We will discuss these somewhat complementary possibilities in turn.

\subsection{Gravitational Waves}\label{g-ws}

\begin{figure}[t]
	\begin{center}
	   \includegraphics[width=0.49\textwidth]{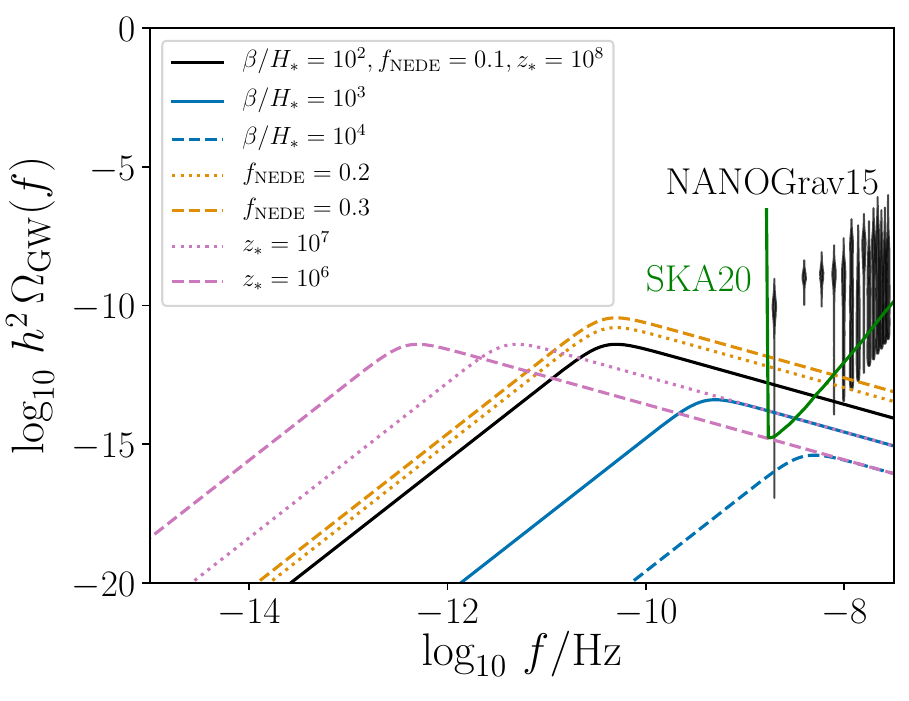}
    \end{center}
    \caption{Gravitational wave spectrum generated in the first-order phase transition using the envelope approximation. We show an estimated prediction for a benchmark scenario (black) and the impact when varying one of the parameters (other lines, see legend). We also include NANOGrav 15yr results from~\cite{NANOGrav:2023gor} and the expected sensitivity of SKA after 20 years of data.} \label{fig:GW}
\end{figure}

A stochastic gravitational wave background is a typical prediction of a cosmological first-order phase transition. In general, there are different production mechanisms, related either to plasma effects such as sound waves and turbulence, or the bubble walls. To keep the discussion simple, we will focus here on bubble collisions as the source of gravitational waves (additional sources might lead to a stronger signal \cite{Jinno:2020eqg,Jinno:2021ury,Jinno:2022mie,Hindmarsh:2013xza,Hindmarsh:2015qta,Hindmarsh:2016lnk,Hindmarsh:2017gnf}). In that case, the spectrum produced has a characteristic peak determined by the typical size of the nucleated bubbles relative to the Hubble scale at the time when they collide (we follow~\cite{Caprini:2018mtu} and \cite{Niedermann:2020dwg})
 \bea\label{gw-spectrum}
 h^2 \Omega_{\rm GW} &=& 6 \times 10^{-8} \left(H_* {\beta}^{-1} \right)^2\nonumber \\ &\times& \left( \frac{f_\text{NEDE}}{0.1} \right)^2 (g_\mathrm{rel,vis}^*)^{-1/3}   S_{\rm GW}(f)\,,
 \eea
where $\beta^{-1}$ sets the time-scale for the duration of the transition, see~\eqref{eq:beta}, and the strength of the transition is commonly characterized by $\alpha = \Delta V_*/(\rho^*_\mathrm{rad,d}+\rho^*_\mathrm{rad,vis})$, which implies $\alpha= f_\mathrm{NEDE}/(1-f_\mathrm{NEDE})$, see~\eqref{fNEDE}. Further,
\begin{align}\label{eq:spectral_shape}
 S_{\rm GW}(f) \simeq \frac{3.8\left( f/f_* \right)^{2.8}}{1 + 2.8 \left( f/f_*\right)^{3.8}}\,,
\end{align}
is the spectral shape obtained in the envelope approximation~\cite{Huber:2008hg} (see also \cite{Caprini:2007xq,Caprini:2009fx,Huber:2008hg,Weir:2016tov,Jinno:2016vai,Jinno:2017fby,Konstandin:2017sat}), and 
\begin{align}\label{f-peak}
f_* \simeq 4.1 \times 10^{-6} \,  \text{nHz} \, \frac{1}{\left(H_* \beta^{-1}\right)} \left(\frac{1+z_\ast}{10^6} \right) \left( g_\mathrm{rel,vis}^* \right)^{1/6}\,,
\end{align}
is the peak frequency as measured today. Before the peak the spectrum grows as $f^3$ for $f \ll f_*$ and falls off after the peak as $f^{-1}$ for $f \gg f_*$.  Even in the extreme case where the phase transition happens just after BBN, with $z_\ast \lesssim  z_\mathrm{BBN} \sim 10^{9}$, the peak frequency is smaller than nano-Hertz. Therefore,  only in the regime $f \gg f_*$,  we can have hope of detecting a signal with pulsar timing arrays (PTA) probing $f \sim \text{nHz}$~\cite{Moore:2014lga}. To find the amplitude of the signal in the nano-Hertz regime, one can substitute \eqref{f-peak} into \eqref{gw-spectrum}
\bea\label{eq:high-f-tail}
  \Omega_{\rm GW}h^2 &\simeq&   3.4 \times 10^{-13} \, \left(\frac{H_* \beta^{-1}}{0.01} \right)\,\left(\frac{ \text{nHz}}{f}\right)\,\nonumber\\
 &\times&\left( \frac{f_\text{NEDE}}{0.1} \right)^2  \left( g_\mathrm{rel,vis}^* \right)^{-1/6}\,\left(\frac{1+z_\ast}{10^8} \right)\,,
\eea
where the ratio $H_*/\beta$, due to \eqref{eq:beta}, is bounded from above by $10^{-2}$ (corresponding to an order unity gauge coupling). For example, given the peak frequency of the \textit{Square Kilometer Array} (SKA) of~\cite{Carilli:2004nx,Janssen:2014dka,Weltman:2018zrl} $\Omega_{\rm GW}h^2 \sim 10^{-15}$, we could expect a signal for a phase transition that occurs in the range $z_\mathrm{BBN}  > z \gtrsim 10^6$. We display in Fig.~\ref{fig:GW} some scenarios for the gravitational wave prediction from the envelope approximation, together with results from 15 years of pulsar observations by the NANOGrav collaboration~\cite{NANOGrav:2023gor} as well as the expected sensitivity of SKA after 20 years of observation.\footnote{For a detailed analysis of a pre-BBN phase transition in a dark sector including BBN constraints on $N_{\rm eff}$ see \cite{Bringmann:2023opz} (as well as \cite{Cruz:2023lnq} for a discussion of how these constraints can be avoided).} 

We stress that the fall-off in~\eqref{eq:spectral_shape} is obtained by using the envelope approximation, which is subject to rather large theoretical uncertainties. In particular, recent lattice calculations indicate that the fall-off lies between~\cite{Cutting:2018tjt,Cutting:2020nla} $f^{-1.4}$ and $f^{-2.2}$, which for a post-BBN phase transition with $z_\ast\lesssim 10^9$, would move a signal below the sensitivity threshold. However, the same studies also find the presence of a ``bump'' in the high-frequency tail. It is caused by oscillations of the scalar field around the true vacuum and can raise the power even above the envelope estimate. Moreover, it is not clear if these studies account for the phenomenology of a supercooled phase transition. We therefore leave a more detailed discussion of these issues for future work and maintain that Fig.~\ref{fig:GW} can still provide us with an indication of where to look for a signal.  

\subsection{Small scales}\label{sec:smallscales}

\begin{figure}[t]
	\centering
	   \includegraphics[width=0.49\textwidth]{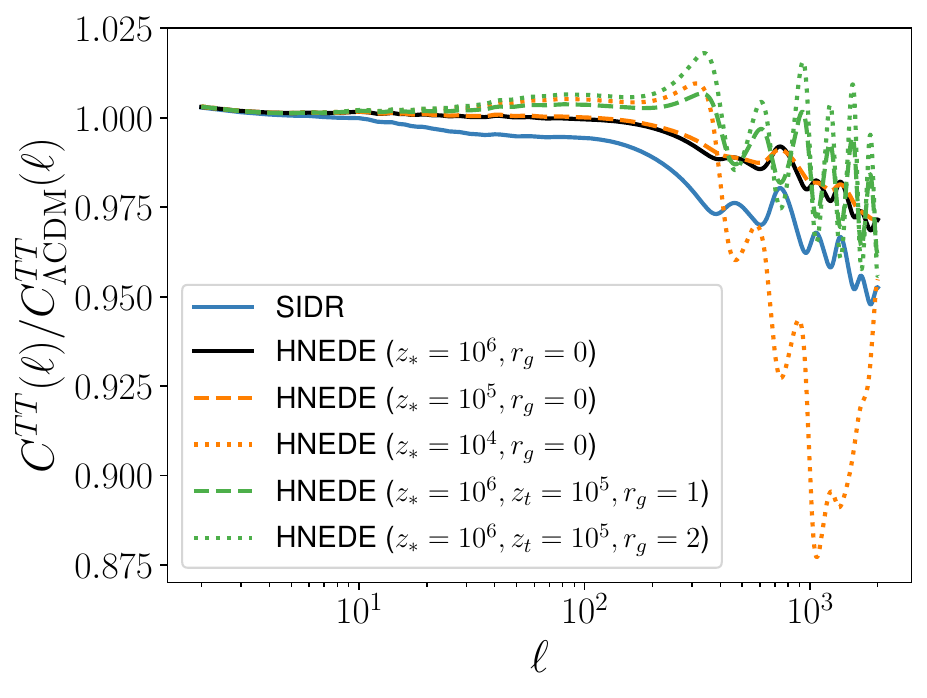}
	   \includegraphics[width=0.49\textwidth]{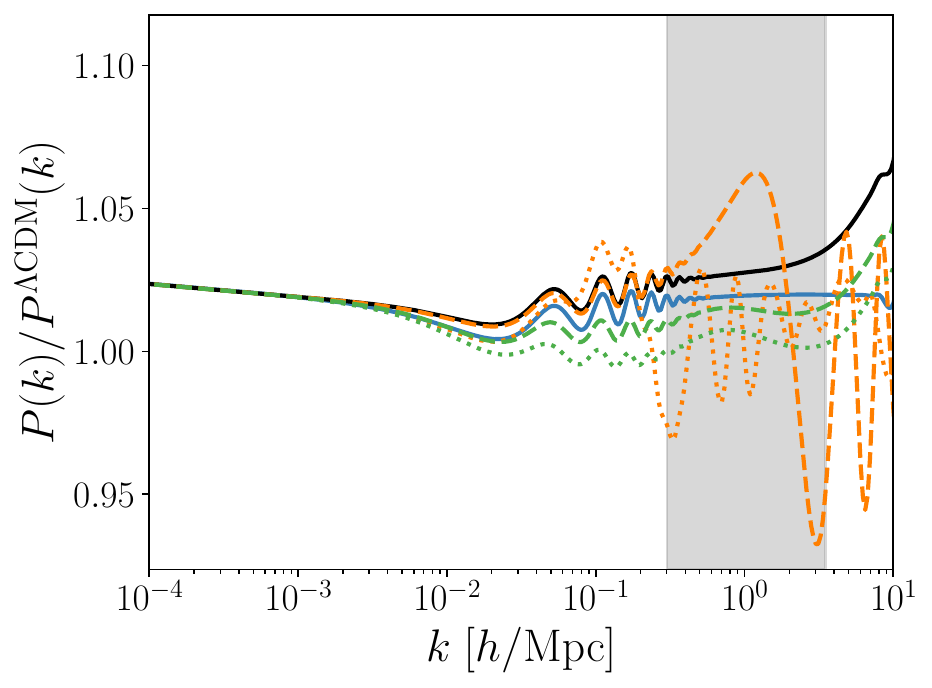}
    \caption{\label{fig:pk} CMB $C_{\ell}^{TT}$ (upper panel) and matter power spectra $P(k)$ (lower panel) for Hot NEDE models with various $z_\ast$ and $r_g$, normalized to $\Lambda$CDM. In all cases, we use the bestfit (base+BBN$+H_0$) parameters obtained for Hot NEDE with $z_\ast = 10^6,\, r_g = 0$ and match $\Delta N_{\rm IR} = 0.40$ for SIDR. Compatibility with Planck requires $z_\ast\gtrsim 10^5$, while for  $z_\ast \lesssim 10^6$ Lyman-$\alpha$ scales overlap with the onset (yellow dashed) or even the first peaks (yellow dotted) of the dark sector acoustic oscillations. The grey area represents the region probed by Lyman-$\alpha$ forest data.  }
\end{figure}

The first-order phase transition imprints characteristic features on the small-scale matter power spectrum, affecting modes that have already entered the horizon at the transition $z_\ast$, i.e. on scales 
\begin{equation}
 k\gtrsim \frac{0.3h}{\rm Mpc}\frac{1+z_\ast}{10^5}\,.
\end{equation}
As discussed in Sec.~\ref{sec:results} and~\ref{subsec:perts} , the adiabatic perturbations present in the dark radiation trigger the phase transition at slightly different times in different locations, and kick off fluctuations on the dark plasma after the transition on sub-horizon scales whose propagation leads to dark acoustic oscillations. Since the dark radiation constitutes a non-negligible fraction of the energy density {\it after} the transition, $\rho_{\rm NEDE}/\rho_{\rm vis}\simeq 8\%\times(\Delta N_{\rm eff}/0.6)$, they impact the gravitational potential and consequently imprint oscillatory features on the CMB as well as on the matter power spectrum, see Fig.~\ref{fig:pk}. For $z_\ast=10^4$ these features overlap with the scales probed by the CMB, excluding these low redshifts, requiring $z_\ast\gtrsim 4\times 10^4$ as discussed in Sec.~\ref{sec:results}. However, for $z_\ast\sim 10^5-10^6$ the dark acoustic oscillations show up on smaller scales $k\gtrsim 0.3h/$Mpc where they can potentially be probed by large-scale structure observations. We indicate the scales probed by one-dimensional Lyman-$\alpha$ forest measurements by the gray band in Fig.~\ref{fig:pk}, for which the ongoing DESI survey~\cite{DESI:2016fyo} is expected to achieve percent-level sensitivity~\cite{Karacayli:2020aad,Walther:2020hxc}, see~\cite{Karacayli:2023afs} for first results from the early data release. We also show the power spectrum for the case with the same amount of DR but without phase transition (SIDR) in Fig.~\ref{fig:pk} for comparison, that is likely indistinguishable from $\Lambda$CDM. Thus, small-scale probes of the power spectrum may discriminate SIDR from the Hot NEDE scenario if the phase transition occurs for $10^6\gtrsim z_\ast\gtrsim 10^5$.

\section{Conclusion and outlook}
\label{sec:conslusion}

In this work we point out a conceptually simple model that can address the Hubble tension by adding extra dark radiation around recombination, while retaining the success of BBN. Its central feature is a supercooled first-order phase transition that occurs in a dark sector between the BBN and recombination epochs. The latent heat released in the phase transition heats up the dark sector, such that the amount of extra radiation can be sizeable during recombination while it is negligibly small during BBN.

We find that all properties of such a dark sector that are favourable from the point of view of cosmology can be realized within a straightforward microscopic model. It features a (dark) gauge symmetry that is spontaneously broken as $SU(N)\to SU(N\!-\!1)$ by a scalar field. The associated phase transition is of first order and features strong supercooling if the scalar is described by a classically (nearly) scale-invariant Lagrangian, leading to radiative symmetry breaking \`a la Coleman-Weinberg in the limit of vanishing effective mass of the scalar field. A soft mass term breaking scale-invariance provides a natural graceful exit mechanism to terminate the supercooling phase and alongside a well-tempered amount of latent heat. 

Moreover, the dark radiation after the transition is comprised of the massless gauge bosons belonging to the remaining $SU(N\!-\!1)$ symmetry, as well as the light dark Higgs boson also characteristic of the Coleman-Weinberg mechanism. The $SU(N\!-\!1)$ gauge interactions naturally make the dark radiation strongly interacting, thus realizing SIDR. Moreover, once the light Higgs becomes non-relativistic somewhat after the phase transition, another slight increase of the amount of extra radiation occurs, as considered in the stepped SIDR scenario. This model thus naturally connects the previously discussed frameworks with Hot NEDE, thereby UV completing (stepped) SIDR to achieve a consistent cosmological model for addressing the Hubble tension while allowing us to also embrace BBN.

To demonstrate the effectiveness of this set-up we implemented the model in a Boltzmann solver and performed an analysis of Planck CMB, as well as BAO and SNe Ia data, considering in addition a prior on the amount of radiation during BBN. This also allows us to consistently use the prediction of the helium fraction from BBN for the subsequent recombination dynamics. We find that for the model considered in this work the Hubble tension can be reduced to the level of $2.8\sigma$, while for SIDR we find $3.9\sigma$ due to the penalty from spoiling BBN (for comparison, our analysis yields $4.3\sigma$ for $\Lambda$CDM). We note that, when ignoring BBN, SIDR performs similarly to the model considered here. 

The supercooled phase transition between BBN and recombination  leads to various signatures. If it occurs shortly after BBN, say for $z_\ast\sim 10^8$, the stochastic gravitational wave background generated by the transition can potentially be seen in pulsar timing arrays, even though a robust prediction would require further work. If the phase transition occurs in the range $10^6\gtrsim z_\ast\gtrsim 10^5$, the impact of dark acoustic oscillations triggered by the transition can be probed with future DESI Lyman-$\alpha$ forest data. We also find a lower bound $z_\ast\gtrsim 4\times 10^4$ from Planck CMB data, since otherwise the dark oscillations would affect the high-$\ell$ modes. We find the Hot NEDE perturbations triggered by the dark plasma to be stronger compared to the case of a scalar-field induced (Cold NEDE) transition for modes that are already sub-horizon at $z_\ast$.

Our work motivates further exploration, including technical improvements, for example related to the thermalization dynamics immediately after the transition and the gravitational wave production and spectrum (including sound waves), as well as phenomenological aspects.
To be concrete, it would be interesting to investigate connections with dark matter, e.g. dark matter production in the phase transition \cite{Freese:2023fcr} or the possibility that dark matter is charged under the dark $SU(N)$, which may also address the $S_8$ tension seen in various weak lensing data sets via scattering of dark matter with dark radiation~\cite{Buen-Abad:2017gxg,Niedermann:2021vgd}, which may lead to further signatures in galaxy clustering~\cite{Joseph:2022jsf,Rubira:2022xhb} or cluster counts~\cite{Mazoun:2023kid}. Another direction could be the dynamics in the Early Universe that is responsible for creating the small initial dark sector temperature prior to the phase transition. We note that even the purely gravitational interactions with the visible sector could be sufficiently strong to provide this initial population~\cite{Garny:2015sjg}, with the $SU(N)$ gauge interactions leading to efficient thermalization within the dark sector~\cite{Garny:2018grs}. 

\bigskip

\begin{acknowledgments}
MG thanks Martin Schmaltz for useful conversation, and HR Carlo Tasillo for support with NANOGrav data.
We acknowledge support by the Excellence Cluster ORIGINS, which is funded by the Deutsche Forschungsgemeinschaft (DFG, German Research Foundation) under Germany’s Excellence Strategy - EXC-2094 - 390783311. The work of F.N. is supported by VR Starting Grant 2022-03160 of the Swedish Research Council. M.S.S. is supported by Independent Research Fund Denmark grant 0135-00378B.
\end{acknowledgments}

\appendix

\section{Results including the second step} \label{app:resultssecondstep}

In this appendix we provide further results when explicitly including the second (small) step in $\Delta N_{\rm eff}$ due to the dark Higgs becoming non-relativistic at some redshift $z_t<z_\ast$. Such a step was first considered in \cite{Aloni:2021eaq} as a solution to the Hubble tension (without considering a phase transition at $z_\ast$), and we review its dynamics in App.\,\ref{app:secondstep}. 

\begin{figure}[t]
	\centering
	\includegraphics[width = 0.5\textwidth]{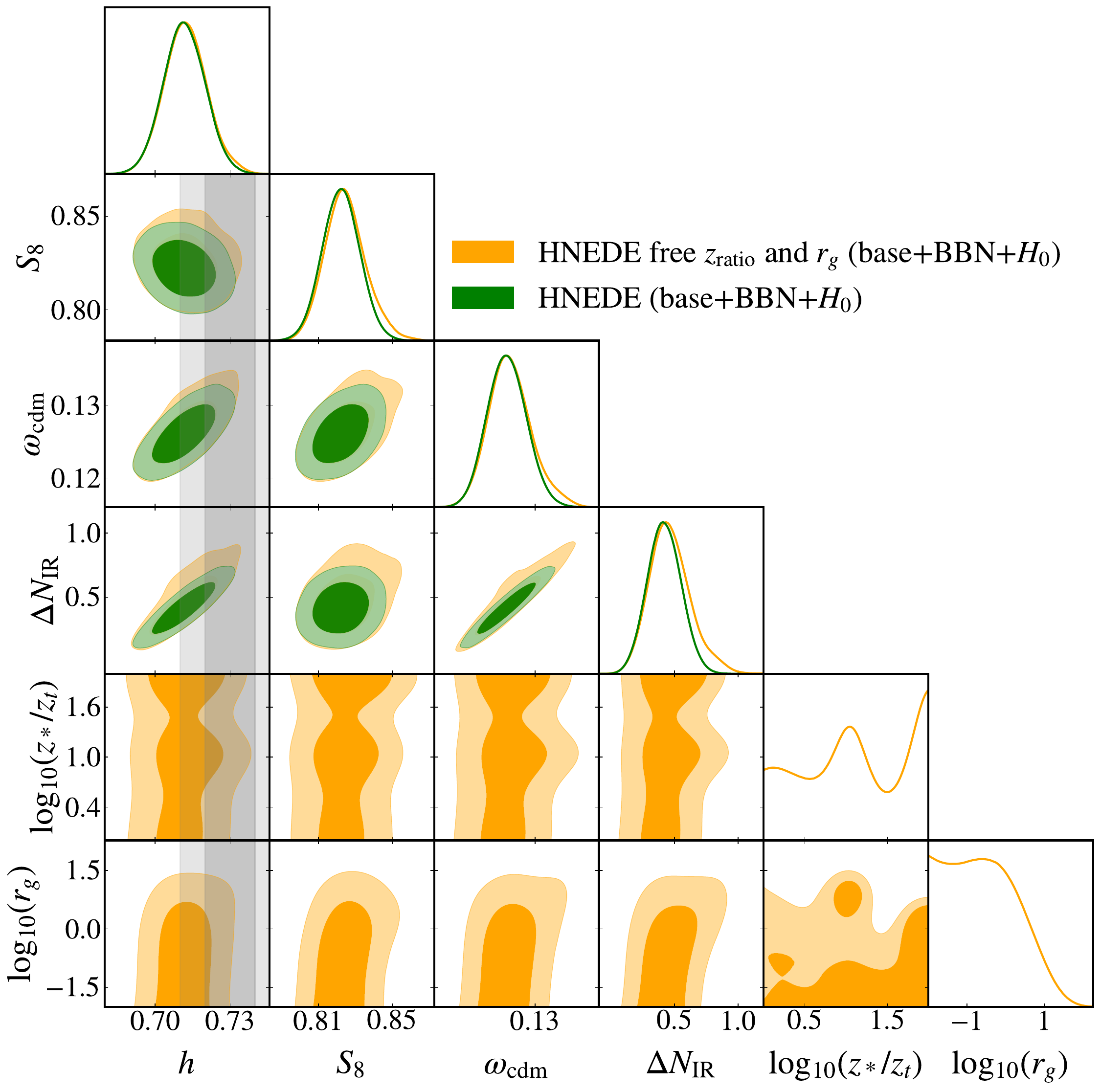}
	\caption{Posteriors exploring the second step at redshift $z_t$ for which the dark Higgs becomes non-relativistic after the phase transition at $z_\ast$ in the Hot NEDE model. We compare the model considered in the main text with step size $r_g = 0$ (green contours) and another one where $r_g$ and $z_t$ are sampled (orange contours). While $r_g$ is bounded from above, $z_t$ is hardly constrained at all, in agreement with findings in the literature for stepped SIDR models~\cite{Allali:2023zbi}.}
	\label{fig:resultsHNEDE}
\end{figure}

We display in Fig.~\ref{fig:resultsHNEDE} the posteriors when including SH0ES data and considering the second step with size $r_g$ (see \ref{def:rg}) at redshift $z_t$, sampled as $\log_{10} r_g$ and $\log_{10} z_\ast/z_t$, respectively (keeping $z_\ast = 10^6$ fixed). We find $\log_{10}(r_g) < 1.03$ ($95\%$ C.L.) and also a relatively unconstrained redshift $z_t$ for the second step. That is consistent with the findings in Fig.~\ref{fig:pk}, in which we notice that the effect of $r_g$ in the angular power spectrum $C^{TT}$ and in the matter power spectrum $P(k)$ is relatively small. Notice that the second step within the $SU(N)$ model is indeed expected to be small with $r_g \leq 1/6$ in agreement with the phenomenological bound. This can be compared with the WZDR model in~\cite{Aloni:2021eaq}, which predicts $r_g^\mathrm{WZDR} = 8/7$. However, this model suffers from BBN constraints similarly to the simplest SIDR model discussed in the main text. 

Note that within the $SU(N)$ model, the redshift of the second step, as derived in~\eqref{def:z_t}, is theoretically constrained to occur within a few $e$-folds after the phase transition, depending on the gauge coupling $g$. In contrast, the step size $r_g$ depends only on the size $N$ of the gauge group. More precise high-$\ell$ CMB data may be able to discriminate scenarios with different step size and redshift, which would therefore allow to specifically constrain $N$ and $g$, respectively.

\section{Review of formalism for the second step} \label{app:secondstep}

In this appendix we review the dynamics of the second step, following~\cite{Aloni:2021eaq}. The expressiond are used for our \texttt{CLASS} implementation of the dynamics around $z_t$ within Hot NEDE.
We consider a dark sector Higgs field with $g_h$ degrees of freedom ($g_h=1$ in our model), mass $m$, temperature $T_d$, and vanishing chemical potential\footnote{The gauge bosons interactions fix $\mu_{\rm gauge} = 0$ and the Higgs gauge boson interactions set $\mu_{h} = 2\mu_{\rm gauge}$.}, and use the dimensionless variable  
\bea
x\left(a(t)\right) = m/T_d\left(a(t)\right)\,.
\eea
We first recall the expressions for the Maxwell-Boltzmann and Bose-Einstein distributions with vanishing chemical potential for a single degree of freedom,
\bea
\rho_{\rm MB}(T_d) &=&  \int_{\bm{q}} \sqrt{q^2+m^2} e^{-\sqrt{q^2+m^2}/T_d}  \vs
&=& \frac{3 T_d^4}{\pi^2} \left( \frac{K_1(x)}{6}x^3 + \frac{K_2(x)}{2}x^2  \right) \,,\nonumber\\
\rho_{\rm BE}(T_d) &=&  \int_{\bm{q}} \sqrt{q^2+m^2} \left[ e^{\sqrt{q^2+m^2}/T_d} - 1 \right]^{-1} \label{eq:rhoBEdef}\,,
\eea
where $\int_{\bm q}\equiv\int\frac{d^3q}{(2\pi)^3}$, and $K_i$ are the modified Bessel functions of the second kind. Similar expressions can be obtained for the pressure. In the high-temperature limit $T_d\gg m$ 
\bea
\rho_{\rm MB}^{\rm high T}(T_d) &=& 
\frac{3}{\pi^2}T_d^4 \,,\nonumber\\
\rho_{\rm BE}^{\rm high T}(T_d) &=& 
\frac{\pi^2}{30}T_d^4 \,,
\eea
with $\rho^{\rm high T}(T_d) = 3 p^{\rm high T}(T_d) $ for both statistics.
We define the (dimensionless) normalized density and pressure for the Maxwell-Boltzmann statistics as 
\bea
\hat{\rho}(T) &\equiv& \rho_{\rm MB}(T) / \rho_{\rm MB}^{\rm high T}(T) = \frac{K_1(x)}{6}x^3 + \frac{K_2(x)}{2}x^2   \,, \nonumber\\
\hat{p}(T) &\equiv& p_{\rm MB}(T) / p_{\rm MB}^{\rm high T}(T) = \frac{K_2(x)}{2}x^2   \,.
\eea
Following~\cite{Aloni:2021eaq}, we  approximate the dark Higgs density and pressure as
\bea
\rho_{h}(T_d) &\equiv& g_h  \rho_{\rm BE}^{\rm high T}  \hat{\rho}_{\rm MB}(T_d)\,,\\
p_{h}(T_d) &\equiv& g_h  p_{\rm BE}^{\rm high T}  \hat{p}_{\rm MB}(T_d)\,.
\eea
Notice that, in the limit $T_d \gg m$, one has $\hat{\rho}(T) \to 1$ and  $\rho_{h}(T) \to  g_h  \rho_{\rm BE}^{\rm high T}$. In other words, the approximation is normalized such that it reproduced the correct quantum statistical result in the high-$T$ limit. It slightly deviates from quantum statistics in the low-$T$ regime~\cite{Aloni:2021eaq}, where the distribution is already Boltzmann-suppressed.  

The total dark sector density and pressure at $z<z_\ast$, including also the massless $SU(N\!-\!1)$ gauge bosons after the phase transition, can be written as
\bea
\rho_d(T_d) &=& \rho_h(T_d) + \frac{\pi^2 g_{\rm gauge} T^4_d}{30}  
\vs
&=& \frac{\pi^2 g_{\rm gauge} T^4_d}{30} \left( 1 + r_g \hat{\rho}(x)\right) \label{eq:rhotorhohat} \,,\\
p_d(T_d) &=& p_h(T_d) + \frac{\pi^2 g_{\rm gauge} T^4_d}{90} 
\vs
&=& \frac{\pi^2 g_{\rm gauge} T^4_d}{90} \left( 1 + r_g \hat{p}(x)\right) \,,
\label{eq:ptophat}
\eea
where $g_{\rm gauge}$ is the number of relativistic degrees of freedom in the gauge sector, with the same dark temperature $T_d$, and
\bea\label{def:rg}
r_g \equiv \frac{g_{h}}{g_{\rm gauge}}\,,
\eea
following the notation used in~\cite{Aloni:2021eaq} for the corresponding quantity.
For the dark sector with $SU(N)\to SU(N\!-\!1)$ symmetry breaking, we have $g_h=1$ and $g_{\rm gauge}=2N(N-2)$, i.e. $1/6\geq r_g\geq 0$ for $3\leq N<\infty$. That leads to a substantially smaller step if compared to the WZDR model of \cite{Aloni:2021eaq}, in which $r_g^{\rm WZDR} = 8/7$.

Parameterizing the energy density in units  of the neutrino energy density $\rho_{1\nu} = \frac{7}{4}\frac{\pi^2}{30}(\frac{T_{\nu 0}}{a})^{4}$ via
\be
\rho_{d}(x) \equiv \Delta N_\mathrm{eff}(x)\rho_{1\nu}\,,
\ee
we find after matching with \eqref{eq:rhotorhohat}
\be
\Delta N_\mathrm{eff}(x) = \Delta N_{\rm IR} \frac{1 + r_g \hat{\rho}(x)}{\left[ 1 + r_g \left( \frac{3}{4}\hat{\rho}(x) + \frac{1}{4}\hat{p}(x)\right) \right]^{4/3}}\,.
\ee
This allows us to identify the effective number of relativistic degrees of freedom before (``NEDE'') and after (``IR'') the step as
\begin{align}
\Delta N_{\rm NEDE}  &= \frac{\Delta N_{\rm IR}}{(1+r_g)^{1/3}} \,,\\
\Delta N_{\rm IR} &= \frac{g_{\rm gauge}}{7/4} \left( \frac{T_{d0}}{T_{\nu 0}} \right)^4 \,.
\end{align}
Using \eqref{eq:rhotorhohat} and \eqref{eq:ptophat}, we obtain for the equation of state and the speed of sound 
\bea 
w(x) &=& \frac{1}{3} - \frac{r_g}{3}\frac{\hat{\rho}(x) - \hat{p}(x)}{1+r_g\hat{\rho}(x)} \,, \label{eq:w_NEDE}\\
c_s^2(x) &=& \frac13-\frac{r_g}{3}\frac{\hat\rho-\hat p-\frac{x}{4}(\hat \rho'-\hat p')}{1+r_g\hat\rho-\frac{x}{4}r_g\hat \rho'}  \,.\label{eq:cs_NEDE}
\eea
In order to derive $x(a)$, needed for the Boltzmann implementation, we use entropy conservation within the dark sector for $z>z_\ast$, implying $x\times (\rho+p)\propto 1/a^3$. This can be written as
\bea
\left(\frac{xa_{t}}{a}\right)^3 = 1 + \frac{r_g}{4}\left( 3 \hat{\rho}(x)+\hat{p}(x)\right) \,,
\eea
and solved for $a$ 
\bea
a = a_t x \left( \frac{1}{1 + \frac{r_g}{4}\left[ 3 \hat{\rho}(x)+\hat{p}(x)\right]}\right)^{1/3} \,,
\eea
where $a_t \equiv 1/(1+z_t)$ is the scale at which the Higgs field becomes non-relativistic as defined in \cite{Aloni:2021eaq} and used as an input parameter for the Boltzmann solver in the form of $z_{\rm ratio} = z_\ast/z_t $. This expression can be inverted numerically to obtain $x(a)$.\footnote{Note that the energy conservation equation $d\rho_d/dt+3H(\rho_d+p_d)=0$ in the dark sector is also automatically satisfied in this case. We can rewrite it as $a\frac{dx}{da}\frac{d\rho_d}{dx}=-3(\rho_d+p_d)$. Writing the entropy equation as $x(\rho_d+p_d)\propto 1/a^3$ and taking a derivative of this equation with respect to $a$, one can check that energy conservation requires that $\rho_d+p_d+x\frac{dp_d}{dx}=0$. This is satisfied for the contribution from gauge bosons. For the Higgs, it requires $\hat \rho-\hat p+\frac{x}{3}\frac{d\hat p}{dx}=0$. Using explicit expressions from above, one sees that this is indeed satisfied.}

\bibliography{ref}

\end{document}